\documentclass{acmconfbig}
\usepackage{graphicx} 
\usepackage{epsfig}
\usepackage{stmaryrd}
\usepackage{latexsym}
\usepackage{amssymb}
\usepackage{daytime}
\usepackage{eepic}
\usepackage{url}
\usepackage{verbatim}
\def\thisPaperTitle{The First-Order Theory of Sets with Cardinality Constraints is Decidable}

\usepackage{defs}
\title{\thisPaperTitle}

\author{Viktor Kuncak and Martin Rinard \\
        Computer Science and Artificial Intelligence Laboratory \\
        Massachusetts Institute of Technology\\
        Cambridge, MA 02139, USA \\
        {\tt $\{$vkuncak,rinard$\}$@csail.mit.edu} \\
        {MIT CSAIL Technical Report 958} \\
        {\tiny VK0120, July 2004}
        \vspace*{-2em}
}

\begin{document}

\sloppy

\maketitle

\renewcommand{\thefootnote}{\fnsymbol{footnote}}
\footnotetext{Version compiled \today, \Daytime.}

\renewcommand{\thefootnote}{\arabic{footnote}}

\begin{abstract}
Data structures often use an integer variable to keep track
of the number of elements they store.  An invariant of such
data structure is that the value of the integer variable is
equal to the number of elements stored in the data
structure.  Using a program analysis framework that supports
abstraction of data structures as sets, such constraints can
be expressed using the language of sets with cardinality
constraints.  The same language can be used to express 
preconditions that guarantee the correct use of the data
structure interfaces, and to express invariants useful for
the analysis of the termination behavior of programs that
manipulate objects stored in data structures.  In this paper
we show the decidability of valid formulas in one such
language.

Specifically, we examine the first-order theory that
combines 1) Boolean algebras of sets of uninterpreted
elements and 2) Presburger arithmetic operations.  Our
language allows relating the cardinalities of sets to the
values of integer variables.  We use quantifier elimination
to show the decidability of the resulting first-order
theory.  We thereby disprove a recent conjecture that this
theory is undecidable.  We describe a basic
quantifier-elimination algorithm and its more sophisticated
versions.  From the analysis of our algorithms we obtain an
elementary upper bound on the complexity of the resulting
combination.  Furthermore, our algorithm yields the
decidability of a combination of sets of uninterpreted
elements with any decidable extension of Presburger
arithmetic.  For example, we obtain decidability of monadic
second-order logic of n-successors extended with sets of
uninterpreted elements and their cardinalities, a result
which is in contrast to the undecidability of extensions of
monadic-second order logic over strings with equicardinality
operator on sets of strings.
\end{abstract}




\section{Introduction}

Program analysis and verification tools can greatly
contribute to software reliability, especially when used
throughout the software development process.  Such tools are
even more valuable if their behavior is predictable, if they
can be applied to partial programs, and if they allow the
developer to communicate the design information in the form of
specifications.  Combining the basic idea of
\cite{Hoare69AxiomaticBasisComputerProgramming} with decidable
logics leads to analysis tools that have these desirable
properties, examples include \cite{Moeller01PALE,
KuncakRinard03OnRoleLogic, BallRajamani00BooleanPrograms,
RepsETAL04SymbolicImplementationBestTransformer,
YorshETAL04SymbolicallyComputingMostPrecise,
LamETAL04GeneralizedTypestateCheckingUsingSets,
LamETAL04ModularPluggableAnalyses}.  These analyses are
precise (because they represent loop-free code
precisely) and predictable (because the checking of
verification conditions terminates either with a realizable
counterexample or with a sound claim that there are no
counterexamples).

The key challenge in this approach to program analysis and
verification is to identify a logic that captures an
interesting class of program properties, but is nevertheless
decidable.  In \cite{LamETAL04ModularPluggableAnalyses,
LamETAL04GeneralizedTypestateCheckingUsingSets} we 
identify the first-order theory of Boolean algebras as a
useful language for languages with dynamically allocated
objects: this language allows expressing generalized
typestate properties and reasoning about data
structures as dynamically changing sets of objects.

The results of this paper are motivated by the fact that we
often need to reason not only about the data structure
content, but also about the size of the data structure.  For
example, we may want to express the fact that the number of
elements stored in a data structure is equal to the value of
an integer variable that is used to cache the data structure
size, or we may want to introduce a decreasing integer
measure on the data structure to show program termination.
These considerations lead to a natural generalization of the
first-order theory of Boolean algebra of sets, a
generalization that allows integer variables in addition to
set variables, and allows stating relations of the form
$|A|=k$ meaning that the cardinality of the set $A$ is equal
to the value of the integer variable $k$.  Once we have
integer variables, a natural question arises: which
relations and operations on integers should we allow?  It
turns out that, using only the Boolean algebra operations
and the cardinality operator, we can already define all
operations of Presburger arithmetic.  This leads to the
structure $\BAPA$, which properly generalizes both Boolean
algebras ($\BA$) and Presburger arithmetic ($\PA$).  Our
paper shows that the first-order theory of structure $\BAPA$
is decidable.

A special case of $\BAPA$ was recently shown decidable in
\cite{Zarba04QuantifierEliminationAlgorithmSetCardinality},
which allows only quantification over \emph{elements} but
not over \emph{sets} of elements.  (Note that quantification
over sets of elements subsumes quantification over elements
because singleton sets can represent elements.)  In fact,
\cite{Zarba04QuantifierEliminationAlgorithmSetCardinality}
identifies the problem of decidability of $\BAPA$ and
conjectures that it is \emph{undecidable}.  Our result
proves this conjecture false by showing that $\BAPA$ is
decidable.  Moreover, we give a translation of $\BAPA$
sentences into $\PA$ sentences and derive an elementary
upper bound on the worst-case complexity of the validity
problem for $\BAPA$.

\smartparagraph{Contributions and Overview.}
We can summarize our paper as follows.
\begin{enumerate}
\item We {\bf motivate} the use of sets with cardinality constraints through
      an example (Section~\ref{sec:example}) and show how to
      reduce the validity of annotated recursive program
      schemas (which are a form of imperative programs) to
      the validity of logic formulas
      (Section~\ref{sec:schemas}).
\item We show the {\bf decidability} of Boolean algebras with Presburger arithmetic
      ($\BAPA$) using quantifier elimination in
      Section~\ref{sec:qeBasic}.  This result immediately
      implies decidability of the verification problem for schemas whose
      specifications are expressed in $\BAPA$.

      As a preparation for this result, we review the
      quantifier elimination technique in
      Section~\ref{sec:qeIntro} and show its application to
      the decidability of Boolean algebras
      (Section~\ref{sec:qeBA}) and Presburger arithmetic
      (Section~\ref{sec:qePresburger}).  We also explain why
      adding the equicardinality operator to Boolean algebras
      allows defining Presburger arithmetic operations on
      equivalence classes of sets (Section~\ref{sec:baequi}).
\item We present an {\bf algorithm} $\alpha$ (Section~\ref{sec:qeReduction})
      that translates $\BAPA$ sentences into $\PA$ sentences
      by translating set quantifiers into integer
      quantifiers.  This is the central result of this paper
      and shows a natural connection between Boolean
      algebras and Presburger arithmetic.
\item We analyze our algorithm $\alpha$ and show that it yields
      an {\bf elementary upper bound} on the worst-case complexity
      of the validity problem for $\BAPA$ sentences that is close
      to the bound on $\PA$ sentences themselves (Section~\ref{sec:complexity}).
\item We show that $\PA$ sentences generated by translating pure
      $\BA$ sentences can be checked for
      validity in the space {\bf optimal for Boolean algebras}
      (Section~\ref{sec:baspecial}).
\item We extend our {\bf algorithm} to {\bf infinite sets}
      and predicates for distinguishing finite and infinite
      sets (Section~\ref{sec:infSets}).
\item We examine the relationship of our results to
      the monadic second-order logic (MSOL) of strings 
      (Section~\ref{sec:msol}).  In contrast to the 
      undecidability of MSOL with equicardinality
      operator (Section~\ref{sec:undmsol}),
      we identify a combination of MSOL over trees with $\BA$ that
      is {\bf decidable}.  This result follows from
      the fact that our algorithm $\alpha$ enables adding
      $\BA$ operations to any extension of Presburger arithmetic,
      including decidable extensions such as MSOL over strings
      (Section~\ref{sec:decmsol}).
\end{enumerate}


\section{Example}
\label{sec:example}

\noindent
Figure~\ref{fig:exampleProcedure} presents a procedure
$\insertd$ in a language that directly manipulates sets.
Such languages can either be directly executed
\cite{Dewar79ProgrammingRefinementSETL, SchonbergETAL91Setl}
or can be derived from executable programs using an
abstraction process
\cite{LamETAL04GeneralizedTypestateCheckingUsingSets,
  LamETAL04ModularPluggableAnalyses}.  The program in
Figure~\ref{fig:exampleProcedure} manipulates a global set
of objects $\content$ and an integer field $\sized$.  The
program maintains an invariant $I$ that the size of the set
$\content$ is equal to the value of the variable $\sized$.
The $\insertd$ procedure inserts an element $e$ into the set
and correspondingly updates the integer variable.  The
requires clause (precondition) of the $\insertd$ procedure
is that the parameter $e$ is a non-null reference to an
object that is not stored in the set $\content$.  The
ensures clause (postcondition) of the procedure is that the
$\sized$ variable after the insertion is positive.  Note
that we represent references to objects (such as the
procedure parameter $e$) as sets with at most one element.
An empty set represents a null reference; a singleton set
$\{o\}$ represents a reference to object $o$.  The value of
a variable after procedure execution is indicated by marking
the variable name with a prime.

\begin{figure}[htbp]
\footnotesize
\begin{equation*}
\begin{array}{l}
\vard\ \content:\set; \\
\vard\ \sized:\integer; \\
\invariant\ I \iff (\sized = |\content|); \\
\ \\
\procedure\ \insertd(e: \element)\ \maintains\ I \\
\requires\ |e|=1 \land |e \cap \content|=0 \\
\ensures\ \sized' > 0\\
\{ \\
\ \ \ \ \content := \content \cup e;\\
\ \ \ \ \sized := \sized + 1;\\
\}
\end{array}
\end{equation*}
\caption{An Example Procedure\label{fig:exampleProcedure}}
\begin{equation*}
\begin{array}{l}
\big\{ |e|=1 \land |e \cap \content|=0 \land \sized=|\content| \big\} \mnl
\ \ \ \ \content := \content \cup e;\ \ \sized := \sized + 1; \mnl
\big\{\sized' > 0 \land \sized' = |\content'| \big\}
\end{array}
\end{equation*}
\caption{Hoare Triple for $\insertd$ Procedure\label{fig:exampleTriple}}
\end{figure}

\begin{figure}[htbp]
\footnotesize
\begin{equation*}
\begin{array}{l}
\forall e.\ \forall \content.\ \forall \content'.\ \forall \sized.\ \forall \sized'.\ \\
\ \ 
\begin{array}{@{}l}
  (|e|=1 \land |e \cap \content|=0 \land \sized=|\content| \ \land \\
  \ \content' = \content \cup e \land \sized' = \sized + 1) \implies \\
  \ \ \ \ \sized' > 0 \land \sized' = |\content'|
\end{array}
\end{array}
\end{equation*}
\caption{Verification Condition for Figure~\ref{fig:exampleTriple}\label{fig:exampleVC}}
\end{figure}

In addition to the explicit requires and ensures clauses,
the $\insertd$ procedure maintains an invariant, $I$, which
captures the relationship between the size of the set
$\content$ and the integer variable $\sized$.  The invariant
$I$ is implicitly conjoined with the requires and the
ensures clause of the procedure.  The Hoare triple
\cite{Floyd67AssigningMeaningsPrograms,
  Hoare69AxiomaticBasisComputerProgramming} in
Figure~\ref{fig:exampleTriple} summarizes the resulting
correctness condition for the $\insertd$ procedure.

Figure~\ref{fig:exampleVC} presents a verification condition
corresponding to the Hoare triple in
Figure~\ref{fig:exampleTriple}.  Note that the verification
condition contains both set and integer variables, contains
quantification over these variables, and relates the sizes
of sets to the values of integer variables.  Our small
example leads to a particularly simple formula; in general,
formulas that arise in compositional analysis of set
programs with integer variables may contain alternations of
existential and universal variables over both integers and
sets.  This paper shows the decidability of such formulas.


\section{First-Order-Logic Program Schemas}
\label{sec:schemas}

To formalize the verification of programs with
specifications written in first-order logic, we introduce
the notion of first-order-logic program schemas (or
\emph{schemas} for short).  The schemas motivate the main
result of this paper because the decidability of a class of
logic formulas implies the decidability of the schema
verification problem.  The abstraction of programs in
general-purpose languages into verifiable schemas can be
used to verify partial correctness of programs, and is a
particular instance of abstract interpretation
\cite{CousotCousot79SystematicDesignProgramAnalysis}.
Program schemas have been studied in the past, with the
focus primarily on purely functional schemas
\cite{AshcroftETAL73DecidablePropertiesMonadicFunctionalSchemas,
Chandra73DecisionProblemsProgramSchemasCommutativeInvertible}.

\begin{figure}
\footnotesize
\begin{equation*}
\begin{array}{rcl}
  F & - & \mbox{first-order formula} \mnl
  s & ::= & F \mid p \mid s \seqc\, s \mid s \ndch\, s \mid \var\ x:T.\ s \mnl
  \spec{p} & ::= & 
    \begin{array}[t]{l}
      \procedure\ p \\
      \requires\ \pre{p} \\
      \ensures\ \post{p} \\
      \mathbf{\{} \body{p} \mathbf{\}} \mnl 
    \end{array} \\
  \schema & ::= & \rep{\var\ x:T} \rep{\spec{}}
\end{array}
\end{equation*}
\caption{Syntax of First-Order Logic Program Schemas\label{fig:schemaSyntax}}
\end{figure}

Figure~\ref{fig:schemaSyntax} presents the syntax of
schemas.  A schema is a collection of annotated recursive
procedures that manipulate global state given by finitely
many variables.  A recursive program schema is parameterized
by a specification language which determines 1) a signature
of the specification language, which is some variant of
first-order logic and 2) the interpretation of the
language, which is some family of multisorted first-order
structures.  The interpretations of types of global and
local variables correspond to the interpretations of sorts
in the multisorted language.  We use the term ``$S$-schema''
for a schema parameterized by a specification language $S$.
The language $S$ is used to encode all basic statements of
the schema and to write requires and ensures clauses. The
only control structures in a schema are sequential
composition ``$\seqc$'', nondeterministic choice
``$\ndch$'', and procedure call (denoted using procedure
name).  For simplicity, procedures in a schema have no
parameters; parameter passing can be simulated using
assignments to global and local variables.

\begin{figure}
\newcommand{\cmt}[1]{\multicolumn{3}{c}{\mbox{{#1}:}}}
\newcommand{\scmt}[1]{\multicolumn{3}{c}{\mbox{{#1}}}}
\footnotesize
\begin{equation*}
\begin{array}{rcl}
  \cmt{the meaning of specifications:} \\
  \spec{p} & = & 
(\pre{p} \implies \post{p}) \mnl
  \cmt{rules for reducing statements to formulas:} \\
  p & \transto & \spec{p} \mnl
  F_1 \seqc\, F_2 & \transto & 
        \exists \xbar_0. (F_1[\xbar' := \xbar_0] \land
                          F_2[\xbar := \xbar_0]) \mnl
  \scmt{$\xbar$ - variables in pre-state} \\
  \scmt{$\xbar'$ - variables in post-state} \mnl
  F_1 \ndch\, F_2 & \transto & F_1 \lor F_2 \mnl
  \var\ x:T.\ F & \transto & \exists x:T.\ F \mnl
  \cmt{correctness condition for $p$:} \mnl
  \multicolumn{3}{c}{
    \begin{array}{l}  
     \forall^{*} (\fbody{p} \implies \spec{p}) \mnl
     \mbox{ where \ \ } 
     \body{p} \transto\limits^{*} \fbody{p}
     \mbox{\ \ using rules above} 
   \end{array}} 
 \end{array}
\end{equation*}
\caption{Rules that Reduce Procedure Body to a Formula\label{fig:schemaRules}}
\end{figure}

Provided that variables in $S$ range over sufficiently complex
data types (such as integers or terms), 
schemas are a Turing-complete language.  Indeed,
the first-order logic can encode assignment statement ($x:=t$ is represented by formula
$x'=t \land \bigwedge_{y \not\equiv x} y'=y$), as
well as $\m{assume}$ statements ($\m{assume}\ F$ is just $F \implies \skips$ 
where $\skips$ is $\bigwedge_y y'=y$); 
nondeterministic choice and assume-statements can encode the $\m{if}$
statements; recursion with assume statements can encode
$\m{while}$ loops.  As a consequence of
Turing-completeness, the verification of schemas
\emph{without} specifications would be undecidable.  Because
we are assuming that procedures are annotated, the
correctness of our recursive program schema reduces to the
validity of a set of formulas in the logic, using standard
technique of assume-guarantee reasoning.  The idea of this
reduction is to replace each call to procedure $p$ with the
specification given by requires and ensures clause of
$p$, as in Figure~\ref{fig:schemaRules}.  After this
replacement, the body of each procedure contains only
sequential composition, basic statements, and
nondeterministic choice.  The remaining rules in
Figure~\ref{fig:schemaRules} then reduce the body of a
procedure to a single formula.\footnote{Note that our formulas encode transition relation
as opposed to weakest precondition, so we use $\lor$ to encode non-deterministic 
choice and $\forall$ for uninitialized variables.}
We check the correctness of
the procedure by checking that the formula corresponding to
the body of the procedure implies the specification of the
procedure.

We conclude that if the validity of first-order formulas in
the language $S$ is decidable, then the verification problem
of an $S$-schema parameterized by those formulas is
decidable.  By considering different languages $S$ whose
first-order theory is decidable, we obtain different
verifiable $S$-schemas.  Example languages whose first-order
theories are decidable are term algebras and their
generalizations
\cite{KuncakRinard03TheoryStructuralSubtyping}, Boolean
algebras of sets
\cite{Loewenheim15UeberMoegligkeitenRelativkalkuel} and
Presburger arithmetic
\cite{Presburger29UeberVollstaendigkeitSystemsAritmethikZahlen}.
In this paper we establish the decidability of the
first-order theory $\BAPA$ that combines the quantified
formulas of Boolean algebras of sets and Presburger
arithmetic.  Our result therefore implies the verifiability
of a new class of schemas, namely $\BAPA$-schemas.

\smartparagraph{Schemas and Boolean programs.} For a fixed
set of predicates, Boolean programs used in predicate abstraction
\cite{BallRajamani00BooleanPrograms,
  BallETAL01AutomaticPredicateAbstraction,
  HenzingerETAL04AbstractionsProofs} can be seen as a
particular form of schemas where the first-order variables
range over finite domains.  The assumption about finiteness
of the domain has important consequences: in the finite
domain case the first-order formulas reduce to quantified
Boolean formulas, the schemas are not Turing-complete but
reduce to pushdown automata, and procedure specifications
are not necessary because finite-state properties can be
checked using context-free reachability.  In this paper we
consider schemas where variables may range over infinite
domains, yet the verification problem in the presence of
specifications is decidable.  The advantage of expressive
program schemas is that they are closer to the
implementation languages, which makes the abstraction of
programs into schemas potentially simpler and more precise.

\smartparagraph{Verification using quantifier-free
  formulas.} Note that the rules in
Figure~\ref{fig:schemaRules} do not introduce quantifier
alternations.  This means that we obtain verifiable
$S$-schemas even if we restrict $S$ to be a quantifier-free
language whose formulas have decidable satisfiability
problem.  The advantage of using languages whose full
first-order theory is decidable is that this approach allows
specifications of procedures to use quantifiers to express
parameterization (via universal quantifier) and information
hiding (via existential quantifier).  Moreover, the quantifier
elimination technique which we use in this paper shows how
to eliminate quantifiers from a formula while preserving its
validity.  This means that, instead of first applying rules
in Figure~\ref{fig:schemaRules} and then applying quantifier
elimination, we may first eliminate all quantifiers from
specifications, and then apply rules in
Figure~\ref{fig:schemaRules} yielding a quantifier-free
formula.  This approach may be more efficient because
the decidability of quantifier-free formulas is
easier to establish \cite{Zarba04CombiningSetsElements,
  Nelson81TechniquesProgramVerification,
  RuessShankar01DeconstructingShostak,
  Zarba04CombinationProblemAutomatedReasoning,
  StumpETAL02CVC}.


\section{Overview of Quantifier Elimination}

For completeness, this section introduces quantifier
elimination; quantifier elimination is the central technique
used in this paper.  After reviewing the basic idea of
quantifier elimination in Section~\ref{sec:qeIntro}, we
explain how to use quantifier elimination to show the
decidability of Boolean algebras in Section~\ref{sec:qeBA}.
We show the decidability of Presburger arithmetic in
Section~\ref{sec:qePresburger}.

\subsection{Quantifier Elimination}
\label{sec:qeIntro}

According to \cite[Page 70,
Lemma~2.7.4]{Hodges93ModelTheory},
to eliminate quantifiers from arbitrary formulas,
it suffices to eliminate
$\exists y$ from formulas of the form
\begin{sequation} \label{eqn:qelimStep}
    \exists y.\ \bigwedge_{0 \leq i<n} \psi_i(\bar x,y)
\end{sequation}
where $\bar x$ is a tuple of variables and $\psi_i(\bar x,y)$
is a literal whose all variables are among $\bar x,y$.  The
reason why eliminating formulas of the
form~(\ref{eqn:qelimStep}) suffices is the following.
Suppose that the formula is in prenex form and consider the
innermost quantifier of a formula.  Let $\phi$ be the
subformula containing the quantifier along with the subformula that
is the scope of that quantifier.  If $\phi$ is of the form $\forall
x.\ \phi_0$ we may replace $\phi$ with $\lnot \exists x.\lnot \phi_0$.  Hence, we
may assume that $\phi$ is of the form $\exists x.\ \phi_0$.  We then
transform $\phi_0$ into disjunctive normal form and use the
fact
\begin{sequation} \label{eqn:existsPropagation}
   \exists x.\ (\phi_2 \lor \phi_3) \iff (\exists x.\ \phi_2) \lor (\exists x.\ \phi_3)
\end{sequation}
We conclude that elimination of quantifiers from formulas of
form~(\ref{eqn:qelimStep}) suffices to eliminate the
innermost quantifier.  By repeatedly eliminating innermost
quantifiers we can eliminate all quantifiers from a formula.

We may also assume that $y$ occurs in every literal $\psi_i$,
otherwise we would place the literal outside the existential
quantifier using the fact
\begin{sequation*}
   \exists y.\ (A \land B)  \iff (\exists y. A) \land B
\end{sequation*}
for $y$ not occurring in $B$.

To eliminate variables we often use the following identity
of theory with equality:
\begin{sequation} \label{eqn:replacement}
   \exists x. x = t \land \phi(x)  \iff  \phi(t)
\end{sequation}

The quantifier elimination procedures we present imply the
decidability of the underlying theories, because the
interpretations of function and relation symbols on some
domain $A$ turn out to be effectively computable functions
and relations on $A$.  Therefore, the truth-value of every
formula without variables is computable.  The quantifier
elimination procedures we present are all effective.  To
determine the truth value of a closed formula $\phi$ on a
given model, it therefore suffices to apply the quantifier
elimination procedure to $\phi$, yielding a quantifier free
formula $\psi$, and then evaluate the truth value of $\psi$.

\subsection{Quantifier Elimination for $\BA$}
\label{sec:qeBA}

This section presents a quantifier elimination procedure for
Boolean algebras of finite sets.  We use the symbols for the
set operations as the language of Boolean algebras.  $b_1
\cap b_2$, $b_1 \cup b_2$, $b_1^c$, $\emptyset$, $\Univ$,
correspond to set intersection, set union, set complement,
empty set, and full set, respectively.  We write $b_1
\subseteq b_2$ for $b_1 \cap b_2 = b_1$, and $b_1
\subset b_2$ for the conjunction $b_1 \subseteq b_2 \ \land\
b_1 \neq b_2$.

\begin{figure}
  \begin{sequation*}
    \begin{array}{rcll}
      F & ::= & A \mid F_1 \land F_2 \mid F_1 \lor F_2 \mid \lnot F \mid
                \exists x.F \mid \forall x.F \mnl
      A & ::= & B_1 = B_2 \mid B_1 \subseteq B_2 \mid \\
        & &    \q{|} B \q{|} = C \ \mid\ \q{|} B \q{|} \geq C \mnl
      B & ::= & x \mid \bazero \mid \baone 
               \mid B_1 \cup B_2 \mid B_1 \cap B_2 \mid B^{c} \mnl
      C & ::= & 0 \mid 1 \mid 2 \mid \ldots
    \end{array}
  \end{sequation*}
  \caption{Formulas of Boolean Algebra ($\BA$)
    \label{fig:BAsyntax}}
\end{figure}

For every nonnegative integer constant $k$ we introduce formulas of the form $|b|
\geq k$ expressing that the set denoted by $b$ has at least
$k$ elements, and formulas of the form $|b| = k$ expressing
that the set denoted by $b$ has exactly $k$ elements.  In
this section, cardinality constraints always relate
cardinality of a set to a \emph{constant} integer.  These
properties are first-order definable within Boolean algebra
itself:
\begin{sequation*}\begin{array}{lcl}
|b| \geq 0     &\equiv& \boolTrue \mnl
|b| \geq k{+}1 &\equiv& \exists x.\ x \subset b \ \land\ 
                                  |x| \geq k \mnl
|b| = k   &\equiv& |b| \geq k \ \land\ \lnot |b| \geq k{+}1
\end{array}\end{sequation*}
We call a language which contains terms $|b|\geq k$ and
$|b|=k$ \emph{the language of Boolean algebras with finite
constant cardinality constraints}.  Figure~\ref{fig:BAsyntax}
summarizes the syntax of this language,
which we denote $\BA$.  Because finite constant cardinality
constraints are first-order definable, the language with
finite constant cardinality constraints has the same 
expressive power as the language of Boolean algebras.  Removing
the restriction that integers are constants is, in fact,
what leads to the generalization from Boolean algebras to
Boolean algebras with Presburger arithmetic in
Section~\ref{sec:bapa}, and is the main topic of this
paper.

\smartparagraph{Preliminary observations.}
Every subset relation $b_1 \subseteq b_2$ is equivalent to
$|b_1 \cap b_2^c|=0$, and every equality $b_1 = b_2$ is
equivalent to a conjunction of two subset relations.  It is
therefore sufficient to consider the first-order formulas
whose only atomic formulas are of the form $|b|=k$ and
$|b|\geq k$.  Furthermore, because $k$ denotes constants, we
can eliminate negative literals as follows:
\begin{sequation} \label{eqn:atleastExact}
\begin{array}{rcl}
\lnot |b|=k      &\iff&  |b|=0 \ \lor \cdots \lor\ 
                         |b|=k{-}1 \ \lor\
                         |b| \geq k{+}1 \mnl
\lnot |b|\geq k  &\iff&  |b|=0 \ \lor \cdots \lor\ |b|=k{-}1
\end{array}
\end{sequation}
Every formula in the language of Boolean algebras can
therefore be written in prenex normal form where the matrix (quantifier-free part)
of the formula is a disjunction of conjunctions of atomic
formulas of the form $|b|=k$ and $|b|\geq k$, with no
negative literals.
If a term $b$ contains at least one operation of
arity one or more, we may assume that the constants
$\emptyset$ and $\Univ$ do not appear in $b$, because
$\emptyset$ and $\Univ$ can be simplified away.
Furthermore, the expression $|\emptyset|$ denotes the
integer zero, so all terms of form $|\emptyset|=k$ or
$|\emptyset|\geq k$ evaluate to $\boolTrue$ or $\boolFalse$.
We can therefore simplify every term $b$ so that
either 1) $b$ contains no occurrences of constants
$\emptyset$ and $\Univ$, or 2) $b \equiv \Univ$.

The following lemma is the main idea behind the quantifier
elimination for both $\BA$ in this section and $\BAPA$ in
Section~\ref{sec:bapa}.
\begin{lemma} \label{lemma:mainelim}
  Let $b_1,\ldots,b_n$ be finite disjoint sets, and
  $l_1,\ldots,l_n,k_1,\ldots,k_n$ be natural numbers.   
  Then the following two statements are equivalent:
  \begin{enumerate}
  \item There exists a finite set $y$ such that
  \begin{sequation} \label{eqn:one}
    \bigwedge_{i=1}^n |b_i \cap y| = k_i \land |b_i \cap y^c|=l_i
  \end{sequation}
  \item \
  \begin{sequation} \label{eqn:two}
    \bigwedge_{i=1}^n |b_i| = k_i + l_i
  \end{sequation}
  \end{enumerate}
  Moreover, the statement continues to hold if for any subset of indices $i$
  the conjunct
  $|b_i \cap y| = k_i$ is replaced by $|b_i \cap y| \geq k_i$ or
  $|b_i \cap y^c|=l_i$ is replaced by $|b_i \cap y^c| \geq l_i$,
  provided that $|b_i| = k_i + l_i$ is replaced by
  $|b_i| \geq k_i + l_i$, as indicated in Figure~\ref{fig:BAQERules}.
\end{lemma}
\begin{proof}
  ($\implies$) Suppose that there exists a set $y$
  satisfying~(\ref{eqn:one}).  Because $b_i \cap y$ and $b_i
  \cap y^c$ are disjoint, $|b_i|=|b_i \cap y|+|b_i \cap
  y^c|$, so $|b_i|=k_i+l_i$ when the conjuncts are $|b_i
  \cap y| = k_i \land |b_i \cap y^c|=l_i$, and $|b_i|\geq
  k_i+l_i$ if any of the original conjuncts have inequality.

  ($\explies$) Suppose that~(\ref{eqn:two}) holds.
  First consider the case of equalities.
  Suppose that $|b_i| = k_i + l_i$ for each of the pairwise
  disjoint sets $b_1,\ldots,b_n$.  For each $b_i$ choose a
  subset $y_i \subseteq b_i$ such that $|y_i|=k_i$.  Because
  $|b_i|=k_i+l_i$, we have $|b_i \cap y_i^c|=l_i$.  Having chosen
  $y_1,\ldots,y_n$, let $y = \bigcup_{i=1}^n y_i$.  For $i
  \neq j$ we have $b_i \cap y_j=\emptyset$ and $b_i \cap
  y_j^c = b_i$, so $b_i \cap y = y_i$ and $b_i \cap y^c =
  b_i \cap y_i^c$.  
  By the choice of $y_i$, we conclude that $y$ is
  the desired set for which~(\ref{eqn:one}) holds.  The case
  of inequalities is analogous: for example, in the case
  $|b_i \cap y| \geq k_i \land |b_i \cap y^c|=l_i$, choose
  $y_i \subseteq b_i$ such that $|y_i|=|b_i|-l_i$.
\end{proof}

\smartparagraph{Quantifier elimination for $\BA$.}
We next describe a quantifier elimination procedure for
$\BA$.  This procedure motivates our algorithm in
Section~\ref{sec:bapa}.

We first transform the formula into prenex normal form and
then repeatedly eliminate the innermost quantifier.  As
argued in Section~\ref{sec:qeIntro}, it suffices to show
that we can eliminate an existential quantifier from any
existentially quantified conjunction of literals.  Consider
therefore an arbitrary existentially quantified conjunction
of literals
\begin{sequation*}
    \exists y.\ \bigwedge_{1 \leq i \leq n} \psi_i(\bar x,y)    
\end{sequation*}
where $\psi_i$ is of the form $|b| = k$ or of the form
$|b|\geq k$.  We assume that $y$ occurs in every formula
$\psi_i$.  It follows that no $\psi_i$ contains $|\emptyset|$ or $|\Univ|$.
Let $x_1,\ldots,x_m,y$ be the set of variables occurring in
formulas $\psi_i$ for $1 \leq i \leq n$.

First consider the more general case $m \geq 1$.  Let for
$i_1,\ldots,i_m \in \{0,1\}$, $s_{i_1\ldots i_m} = x_1^{i_1} \cap \cdots \cap x_m^{i_m}$
where $x^0 = x^c$ and $x^1 = x$.
The terms in the set
\begin{sequation*}
   P = \{ s_{i_1\ldots i_m} \mid
           i_1,\ldots,i_m \in \{0,1\} \}
\end{sequation*}
form a partition.  Moreover, every Boolean
algebra term whose variables are among $x_i$ can be
written as a disjoint union of some elements of the
partition $P$.  Any Boolean algebra term containing
$y$ can be written, for some $p,q \geq 0$ as
\begin{sequation*}
\begin{array}{l}
    (u_1 \cap y) \cup \cdots \cup (u_p \cap y) \cup \mnl
    (t_1 \cap y^c) \cup \cdots \cup (t_q \cap y^c)
\end{array}
\end{sequation*}
where $u_1,\ldots,u_p \in P$ are pairwise distinct elements
from the partition and $t_1,\ldots,t_q \in P$ are pairwise
distinct elements from the partition.  Because
\begin{sequation*}
\begin{array}{l}
    |(u_1 \cap y) \cup \cdots \cup (u_p \cap y) \cup
    (t_1 \cap y^c) \cup \cdots \cup (t_q \cap y^c)| = \mnl
\qquad
    |u_1 \cap y| + \cdots +|u_p \cap y| +
    |t_1 \cap y^c| + \cdots + |t_q \cap y^c|
\end{array}
\end{sequation*}
a formula of the form $|b|=k$ can be written as
\begin{sequation*}
   \bigvee_{k_1,\ldots,k_p,l_1,\ldots,l_q}
   \begin{array}[t]{l}
   (|u_1 \cap y|=k_1 \land \cdots \land |u_p \cap y|=k_p \ \land \mnl
   \ |t_1 \cap y^c|=l_1 \land \cdots \land |t_q\cap y^c|=l_p)
   \end{array}
\end{sequation*}
where the disjunction ranges over nonnegative integers
$k_1,\ldots,k_p,l_1,\ldots,l_q \geq 0$ that satisfy
\begin{sequation} \label{eqn:decomposition}
   k_1+\cdots+k_p +l_1+\cdots+l_q=k
\end{sequation}
From~(\ref{eqn:atleastExact}) it follows that we can perform
a similar transformation for formulas of form $|b|\geq k$
(by representing $|b| \geq k$ as boolean combination of
$|b|=k$ formulas, applying~(\ref{eqn:decomposition}), and
traslating the result back into $|b| \geq k$ formulas).
After performing this transformation, we bring the formula
into disjunctive normal form and continue eliminating the
existential quantifier separately for each disjunct, as
argued in Section~\ref{sec:qeIntro}.  We may therefore
assume that all conjuncts $\psi_i$ are of one of the forms:
$|s \cap y|=k$, $|s \cap y^c|=k$, $|s \cap y|\geq k$, and
$|s \cap y^c| \geq k$ where $s \in P$.

\begin{figure}
\footnotesize
\begin{center}
\begin{sequation*}\begin{array}{c|c}
\mbox{original formula} & \mbox{eliminated form} \\ 
\hline
\exists y.\ \ldots \
|b \cap y| \geq k \land |b \cap y^c| \geq l \ \ldots
& |b| \geq k+l \mnl
\exists y.\ \ldots \
|b \cap y| = k \land |b \cap y^c| \geq l \ \ldots
& |b| \geq k+l \mnl
\exists y.\ \ldots \
|b \cap y| \geq k \land |b \cap y^c| = l \ \ldots
& |b| \geq k+l \mnl
\exists y.\ \ldots \
|b \cap y| = k \land |b \cap y^c| = l \ \ldots
& |b| = k+l \mnl
\end{array}\end{sequation*}
\end{center}
\caption{Rules for Eliminating Quantifiers from Boolean Algebra Expressions
  \label{fig:BAQERules}}
\end{figure}

If there are two conjuncts both of which contain $|s \cap
y|$ for the same $s$, then either they are contradictory or
one implies the other.  We therefore assume that for any $s
\in P$, there is at most one conjunct $\psi_i$ containing
$|s \cap y|$.  For analogous reasons we assume that for
every $s \in P$ there is at most one conjunct $\psi_i$
containing $|s \cap y^c|$.  The result of eliminating the
variable $y$ is then given in Figure~\ref{fig:BAQERules}.
These rules are applied for all distinct partitions $s$ for
which $|s \cap y|$ or $|s \cap y^c|$ occurs.
The case when one of the literals containing $|s \cap y|$
does not occur is covered by the case $|s \cap y|\geq k$ for
$k=0$, similarly for a literal containing $|s \cap y^c|$.

It remains to consider the case $m=0$.  Then $y$ is the only
variable occurring in conjuncts $\psi_i$.  Every cardinality
expression $t$ containing only $y$ reduces to one of $|y|$
or $|y^c|$.  If there are multiple literals containing
$|y|$, they are either contradictory or one implies the
others.  We may therefore assume there is at most one
literal containing $|y|$ and at most one literal containing
$|y^c|$.  We eliminate quantifier by applying rules in
Figure~\ref{fig:BAQERules} putting formally $b=\Univ$,
yielding quantifier-free cardinality constraint of the form
$|\Univ|=k$ or of the form $|\Univ|\geq k$, which does not
contain the variable $y$.

This completes the description of quantifier elimination
from an existentially quantified conjunction.  By repeating
this process for all quantifiers we arrive at a
quantifier-free formula $\psi$.  Hence, we have the
following fact.

\begin{fact} 
For every first-order formula $\phi$ in the language of
Boolean algebras with finite cardinality constraints there
exists a quantifier-free formula $\psi$ such that $\psi$ is
a disjunction of conjunctions of literals of form $|b|\geq
k$ and $|b|=k$ (for $k$ denoting constant non-negative
integers) where $b$ are terms of Boolean algebra, the free
variables of $\psi$ are a subset of the free variables of
$\phi$, and $\psi$ is equivalent to $\phi$ on all Boolean
algebras of finite sets.
\end{fact}


\section{First-Order Theory of $\BAPA$ is Decidable}
\label{sec:bapa}

This section presents the main result of this paper: the
first-order theory of Boolean algebras with Presburger
arithmetic ($\BAPA$) is decidable.  We first motivate the
operations of the structure $\BAPA$ in
Section~\ref{sec:baequi}.  We prove the decidability of
$\BAPA$ in Section~\ref{sec:qeBasic} using a quantifier
elimination algorithm that interleaves quantifier
elimination for the Boolean algebra part with quantifier
elimination for the Presburger arithmetic part.  In
Section~\ref{sec:qeReduction} we present another algorithm
($\alpha$) for deciding $\BAPA$, based on the replacement of
set quantifiers with integer quantifiers.  The analysis of
the algorithm $\alpha$ is the subject of
Section~\ref{sec:complexity}, which derives a worst-case
complexity bound on the validity problem for $\BAPA$.

In this section, we interpret Boolean algebras over the
family of all powersets of finite sets.  Our quantifier
elimination is uniform with respect to the size of the
universal set.  Section~\ref{sec:infSets} extends the result
to allow infinite universal sets and reasoning about
finiteness of sets.

\begin{figure}
  \begin{sequation*}
    \begin{array}{rcll}
      F & ::= & A \mid F_1 \land F_2 \mid F_1 \lor F_2 \mid \lnot F \mid \mnl
        & &     \exists x.F \mid \forall x.F \mid
                \exists k.F \mid \forall k.F \mnl
      A & ::= & B_1 = B_2 \mid B_1 \subseteq B_2 \mid \mnls
             && T_1 = T_2 \mid T_1 < T_2 \mid \prdiv{C}{T} \mnl
      B & ::= & x \mid \bazero \mid \baone 
                \mid B_1 \cup B_2 \mid B_1 \cap B_2 \mid B^{c} \mnl
      T & ::= & k \mid C \mid \MAXCARD \mid T_1 + T_2 \mid T_1 - T_2 \mid C \cdot T \mid \ \q{|} B \q{|} \mnl
      C & ::= & \ldots {-2} \mid -1 \mid 0 \mid 1 \mid 2 \ldots
    \end{array}
  \end{sequation*}
  \caption{Formulas of Boolean Algebras with Presburger
    Arithmetic ($\BAPA$) \label{fig:BAPresburgerSyntax}}
\end{figure}

\subsection{From Equicardinality to $\PA$}
\label{sec:baequi}

To motivate the extension of Boolean algebra with all
operations of Presburger arithmetic, we derive these
operations from a single construct: the equicardinality of
sets.

Define the equicardinality relation $\eqcard{b}{b'}$
to hold iff $|b|=|b'|$, and consider $\BA$ extended with
relation $\eqcard{b}{b'}$.  Define the ternary
relation $\pls{b}{b_1}{b_2} \iff (|b|=|b_1|+|b_2|)$ by the formula
\begin{sequation*}
 \exists x_1.\ \exists x_2.\ 
 \begin{array}[t]{l}
    x_1 \cap x_2 = \emptyset\ \land \ b = x_1 \cup x_2 \ \land \\ 
    \eqcard{x_1}{b_1} \land \eqcard{x_2}{b_2}
  \end{array}
\end{sequation*}
The relation $\pls{b}{b_1}{b_2}$ allows us to express addition
using arbitrary sets as representatives for natural numbers.
Moreover, we can represent integers as equivalence classes
of pairs of natural numbers under the equivalence relation 
$(x,y) \sim (u,v) \iff x+v=u+y$.
This construction allows us to express the unary predicate of
being non-negative.  The quantification over pairs of sets
represents quantification over integers, and 
quantification over integers with the addition operation and
the predicate ``being non-negative'' can express all
operations in Figure~\ref{fig:PresburgerSyntax}.

This leads to our formulation of the language $\BAPA$ in
Figure~\ref{fig:BAPresburgerSyntax}, which contains both the
sets and the integers themselves.  Note the language has two
kinds of quantifiers: quantifiers over integers and
quantifiers over sets; we distinguish between these two kinds
by denoting integer variables with symbols such as $k, l$
and set variables with symbols such as $x, y$.  We use the
shorthand $\existsp k.F(k)$ to denote $\exists k. k \geq 0
\land F(k)$ and, similarly $\forallp k.F(k)$ to denote
$\forall k. k \geq 0 \implies F(k)$.  Note that the language
in Figure~\ref{fig:BAPresburgerSyntax} subsumes the language
in Figure~\ref{fig:PresburgerSyntax}.  Furthermore, the
language in Figure~\ref{fig:BAPresburgerSyntax} contains the
formulas of the form $|b|=k$ whose Boolean combinations can
encode all atomic formulas in Figure~\ref{fig:BAsyntax}, as
in Section~\ref{sec:qeBA}.  This implies that the language
in Figure~\ref{fig:BAPresburgerSyntax} properly generalizes
both the language in Figure~\ref{fig:PresburgerSyntax} and
the language in Figure~\ref{fig:BAsyntax}.  Finally, we note that the
$\MAXCARD$ constant denotes the size of the finite universe,
so we require $\MAXCARD=|\Univ|$ (see Section~\ref{sec:infSets} 
for infinite universe case).

\subsection{Basic Algorithm}
\label{sec:qeBasic}

We first present a simple quantifier-elimination algorithm
for $\BAPA$.  As explained in Section~\ref{sec:qeIntro}, it suffices to
eliminate an existential quantifier from a conjunction $F$ of
literals of Figure~\ref{fig:BAPresburgerSyntax}.  We need to
show how to eliminate an integer existential quantifier, and
how to eliminate a set existential quantifier.
By Section~\ref{sec:qeBA}, assume that all occurrences
of set expressions $b$ are within expressions of the form
$|b|$.  Introduce an integer variable $k_i$ for each such expression
$|b_i|$, and write $F$ in the form
\begin{sequation} \label{eqn:separated}
  F \ \equiv\
  \existsp k_1,\ldots,k_p.\ \bigwedge_{i=1}^p |b_i|=k_i \land F_1(k_1,\ldots,k_p)
\end{sequation}
where $F_1$ is a $\PA$ formula.

To eliminate an existential integer quantifier $\exists k$
from the formula $\exists k.F$, observe that $\exists
k.F(k)$ is equivalent to
\begin{sequation*}
  \existsp k_1,\ldots,k_p.\ \bigwedge_{i=1}^p |b_i|=k_i \land  
  \exists k. F_1(k,k_1,\ldots,k_p)
\end{sequation*}
because $k$ does not occur in the first part of the formula.
Using quantifier elimination for Presburger arithmetic,
eliminate $\exists k$ from $\exists k.F_1$ yielding a
quantifier-free formula $F_2(k_1,\ldots,k_m)$.  The formula
$\exists k.F(k)$ is then equivalent to
$F_2(|b_1|,\ldots,|b_m|)$ and the quantifier has been
eliminated.

To eliminate an existential set quantifier $\exists y$ from
the formula $\exists y.F$, proceed as follows.  Start again
from~(\ref{eqn:separated}), and split each $|b_i|$ into sums
of partitions as in Section~\ref{sec:qeBA}.  Specifically,
let $x_1,\ldots,x_n$ where $y \in \{x_1,\ldots,x_n\}$ be all
free set variables in $b_1,\ldots,b_p$, and let
$s_1,\ldots,s_m$ for $m=2^n$ be all set expressions of the form
$\bigcap_{j=1}^n x_j^{\alpha_j}$ for $\alpha_1,\ldots,\alpha_n \in \{0,1\}$.  
Every expression of the form
$|b|$ is equal to an expression of the form
$\sum_{j=1}^q|s_{i_j}|$ for some $i_1,\ldots,i_q$.
Introduce an integer variable $l_i$ for each $|s_i|$ where
$1 \leq i \leq m$, and write $F$ in the form
\begin{sequation} \label{eqn:extractedSplit}
 \begin{array}{l}
  \existsp l_1,\ldots,l_m.\ \existsp k_1,\ldots,k_p.\ \\
    \quad
        \bigwedge_{i=1}^m |s_i|=l_i \ \land\
        \bigwedge_{i=1}^p t_i=k_i \ \land\
  F_1(k_1,\ldots,k_p)
 \end{array}
\end{sequation}
where each $t_i$ is of the form $\sum_{j=1}^q l_{i_j}$ for
some $q$ and some $i_1,\ldots,i_q$ specific to $t_i$.
Note that only the part $\bigwedge_{i=1}^m |s_i|=l_i$
contains set variables, so $\exists y.F$ is equivalent to
\begin{sequation}
 \begin{array}{l}
  \existsp l_1,\ldots,l_m.\ \existsp k_1,\ldots,k_p.\\
   \quad (\exists y. \bigwedge_{i=1}^m |s_i|=l_i) \ \land\
    \bigwedge_{i=1}^p t_i=k_i \ \land\ F_1(k_1,\ldots,k_p)
 \end{array}
\end{sequation}
Next, group each $s_i$ of the form $|s \cap y|$ with the
corresponding $|s \cap y^c|$ and apply
Lemma~\ref{lemma:mainelim} to replace each pair $|s \cap y|=
l_a \land |s \cap y^c|=l_b$ with $|s| = l_a + l_b$.  As a
result, $\exists y. \bigwedge_{i=1}^m |s_i|=l_i$ is replaced
by a quantifier-free formula of the form
$\bigwedge_{i=1}^{m/2} |s'_i|=l_{a_i}+l_{b_i}$.
The entire resulting formula is
\begin{sequation*}
\begin{array}{l}
  \existsp l_1,\ldots,l_m.\ \existsp k_1,\ldots,k_p.\\
   \quad  \bigwedge_{i=1}^{m/2} |s'_i|=l_{a_i}+l_{b_i} \ \land\
        \bigwedge_{i=1}^p t_i=k_i \ \land\
  F_1(k_1,\ldots,k_p)
\end{array}
\end{sequation*}
and contains no set quantifiers, but contains existential integer
quantifiers.  We have already seen how to eliminate
existential integer quantifiers; by repeating the
elimination for each of $l_1,\ldots,l_m,k_1,\ldots,k_p$, we
obtain a quantifier-free formula.  (We can trivially
eliminate each $k_i$ by replacing it with $t_i$, but it
remains to eliminate the exponentially many variables
$l_1,\ldots,l_m$.)

This completes the description of the basic quantifier
elimination algorithm.  This quantifier-elimination
algorithm is a decision procedure for formulas in
Figure~\ref{fig:BAPresburgerSyntax}.  We have therefore
established the decidability of the language $\BAPA$ that
combines Boolean algebras and Presburger arithmetic, solving
the question left open in
\cite{Zarba04QuantifierEliminationAlgorithmSetCardinality}
for the finite universe case.

\begin{theorem}
  The validity of $\BAPA$ sentences over the family of all
  models with finite universe of uninterpreted elements is
  decidable.
\end{theorem}

\smartparagraph{Comparison with Quantifier Elimination for
$\BA$.}  Note the difference in the use of
Lemma~\ref{lemma:mainelim} in the quantifier elimination for
$\BA$ in Section~\ref{sec:qeBA} compared to the use of
Lemma~\ref{lemma:mainelim} in this section:
Section~\ref{sec:qeBA} uses the statement of the lemma when
the cardinalities of sets are known constants, whereas this
section uses the statement of the lemma in a more general
way, creating the appropriate symbolic sum expression for
the cardinality of the resulting sets.  On the other hand,
the algorithm in this section does not need to consider the
case of inequalities for cardinality constraints, because
the handling of negations of cardinality constraints is
hidden in the subsequent quantifier elimination of integer
variables.  This simplification indicates that the
first-order theories $\BA$ and $\PA$ naturally fit together;
the algorithm in Section~\ref{sec:qeReduction} further
supports this impression.

\subsection{Reducing the Number of Introduced Integer Variables}
\label{sec:qeDemand}

This section presents two observations that may reduce the
number of integer variables introduced in the elimination of
set quantifier in Section~\ref{sec:qeBasic}.
The algorithm in Section~\ref{sec:qeBasic} introduces $2^n$
integer variables where $n$ is the number of set variables
in the formula $F$ of~(\ref{eqn:separated}).

First, we observe that it suffices to eliminate the quantifier
$\exists y$ from the conjunction of the conjuncts
$|b_i|=k_i$ where $y$ occurs in $b_i$.  Let
$a_1(y),\ldots,a_q(y)$ be those terms among $b_1,\ldots,b_p$
that contain $y$, and let $x_1,\ldots,x_{n_1}$ be the free
variables in $a_1(y),\ldots,a_q(y)$.  Then it suffices to
introduce $2^{n_1}$ integer variables corresponding to the
the partitions with respect to $x_1,\ldots,x_{n_1}$, which
may be an improvement because $n_1 \leq n$.

The second observation is useful if the number $q$ of terms
$a_1(y),\ldots,a_q(y)$ satisfies the property $2q+1 < n_1$,
i.e.\ there is a large number of variables, but a small
number of terms containing them.  In this case, consider all 
Boolean combinations $t_1,\ldots,t_u$ of the $2q$ expressions
$a_1(\emptyset), a_1(\Univ), a_2(\emptyset),
a_2(\Univ),\ldots, a_q(\emptyset), a_q(\Univ)$.  For each
$a_i$, we have
\begin{sequation*}
  a_i(y) = (y \cap a_i(\emptyset)) \cup
           (y^c \cap a_i(\Univ))
\end{sequation*}
Each $a_i(\emptyset)$ and each $a_i(\Univ)$ is a disjoint
union of the Boolean combinations of
$t_1,\ldots,t_u$, so each $a_i(y)$ is a disjoint union of
Boolean combinations of $y$ and the expressions
$t_1,\ldots,t_u$ that do not contain $y$.  It therefore
suffices to introduce $2^{2q+1}$ integer variables denoting
all terms of the form $y \cap t_i$ and $y^c \cap t_i$, as
opposed to $2^{n_1}$ integer variables.

\subsection{Reduction to Quantified $\PA$ Sentences}
\label{sec:qeReduction}

This section presents an algorithm, denoted $\alpha$, which
reduces a $\BAPA$ sentence to an equivalent $\PA$ sentence
with the same number of quantifier alternations and an
exponential increase in the total size of the formula.
Although we have already established the decidability of
$\BAPA$ in Section~\ref{sec:qeBasic}, the algorithm $\alpha$
of this section is important for several reasons.
\begin{enumerate}
\item Given the space and time bounds for Presburger
arithmetic sentences
\cite{ReddyLoveland78PresburgerBoundedAlternation}, the
algorithm $\alpha$ yields reasonable space and time bounds
for $\BAPA$ sentences.
\item Unlike the algorithm in
Section~\ref{sec:qeBasic}, the algorithm $\alpha$ does not
perform any elimination of integer variables, but instead
produces an equivalent quantified $\PA$ formula.  The
resulting $\PA$ formula can be decided using any decision
procedure for $\PA$, including the decision procedures
based on automata and model-checking
\cite{KlarlundETAL00MONA, 
GaneshETAL02DecidingPresburgerArithmeticModelChecking}.
\item The algorithm $\alpha$ can eliminate set
quantifiers from any extension of Presburger arithmetic.  We
thus obtain a technique for adding a particular form of set
reasoning to every extension of Presburger arithmetic, and
the technique preserves the decidability of the extension.
An example extension where our construction applies is
second-order linear arithmetic i.e.\ monadic second-order
logic of one successors, as well monadic second order logic
of $n$-successors, as we note in Section~\ref{sec:msol}.
\end{enumerate}
We next describe the algorithm $\alpha$ for transforming a
$\BAPA$ sentence $F_0$ into a $\PA$ sentence.  The algorithm
$\alpha$ is similar to the algorithm in
Section~\ref{sec:qeBasic}, but, instead of eliminating the
integer quantifiers, it accumulates them in a $\PA$ formula.

As the first step of the algorithm, transform $F_0$ into
prenex form
\vspace*{-1em}
\begin{sequation}
  Q_p v_p. \ldots Q_1 v_1.\ F(v_1,\ldots,v_p)
\end{sequation}
where $F$ is quantifier-free, and each quantifier $Q_i v_i$
is of one the forms $\exists k$, $\forall k$, $\exists y$,
$\forall y$ where $k$ denotes an integer variable and $y$
denotes a set variable.  As in Section~\ref{sec:qeBasic},
separate $F$ into the set part and the purely Presburger
arithmetic part by expressing all set relations in terms of
$|b|$ terms and by naming each $|b|$, obtaining a formula
of the form~(\ref{eqn:separated}).  Next, split all sets
into disjoint union of cubes $s_1,\ldots,s_m$ for $m=2^n$
where $n$ is the number of all set variables, obtaining
a formula of the form $Q_p v_p. \ldots Q_1 v_1. F$ where $F$
is of the form~(\ref{eqn:extractedSplit}).
Letting $G_1 = F_1(t_1,\ldots,t_p)$, we obtain a formula of the form
\vspace*{-0.5em}
\begin{sequation} \label{eqn:quantifiedReductionInit}
 \begin{array}{l}
  Q_p v_p. \ldots Q_1 v_1.\\
   \qquad
   \existsp l_1,\ldots,l_m.\
        \bigwedge_{i=1}^m |s_i|=l_i \ \land\ G_1
 \end{array}
\end{sequation}
where $G_1$ is a $\PA$ formula and $m=2^n$.
Formula~(\ref{eqn:quantifiedReductionInit}) is the starting
point of the main phase of algorithm $\alpha$.  The main phase of
the algorithm successively eliminates quantifiers
$Q_1 v_1,\ldots,Q_p v_p$ while maintaining a formula of the
form
\vspace*{-0.5em}
\begin{sequation} \label{eqn:quantifiedReduction}
 \begin{array}{l}
  Q_p v_p \ldots Q_r v_r. \\
   \qquad
   \existsp l_1 \ldots l_q.\
        \bigwedge_{i=1}^q |s_i|=l_i \ \land\ G_r
 \end{array}
\end{sequation}
where $G_r$ is a $\PA$ formula, $r$ grows from $1$ to $p+1$,
and $q = 2^e$ where $e$ for $0 \leq e \leq n$ is the number
of set variables among $v_p,\ldots,v_r$.  The list 
$s_1,\ldots,s_q$ is the list of all $2^e$ partitions formed
from the set variables among $v_p,\ldots,v_r$.

We next show how to eliminate the innermost quantifier $Q_r
v_r$ from the formula~(\ref{eqn:quantifiedReduction}).
During this process, the algorithm replaces the formula
$G_r$ with a formula $G_{r+1}$ which has more integer
quantifiers.  If $v_r$ is an integer variable then the
number of sets $q$ remains the same, and if $v_r$ is a set
variable, then $q$ reduces from $2^e$ to $2^{e-1}$.  We next
consider each of the four possibilities $\exists k$,
$\forall k$, $\exists y$, $\forall y$ for the quantifier
$Q_r v_r$.

Consider first the case $\exists k$.  Because $k$ does not
occur in $\bigwedge_{i=1}^q |s_i|=l_i$, simply move the
existential quantifier to $G_r$ and let $G_{r+1} = \exists
k.G_r$, which completes the step.

For universal quantifiers, observe that
\begin{sequation*}
\lnot (
   \existsp l_1 \ldots l_q.\
        \bigwedge_{i=1}^q |s_i|=l_i \ \land\ G_r)
\end{sequation*}
is equivalent to
\begin{sequation*}
   \existsp l_1 \ldots l_q.\
        \bigwedge_{i=1}^q |s_i|=l_i \ \land\ \lnot G_r
\end{sequation*}
because the existential quantifier is used as a let-binding, so
we may first substitute all values $l_i$ into $G_r$, then
perform the negation, and then extract back the definitions
of all values $l_i$.  Given that the universal quantifier
$\forall k$ can be represented as a sequence of unary
operators $\lnot \exists k \lnot$, from the elimination of
$\exists k$ we immediately obtain the elimination of
$\forall k$; it turns out that it suffices to let $G_{r+1} = \forall k.G_r$.

We next show how to eliminate an existential set quantifier
$\exists y$ from 
\begin{sequation} \label{eqn:reduceStart}
   \exists y.\
   \existsp l_1 \ldots l_q.\
        \bigwedge_{i=1}^q |s_i|=l_i \ \land\ G_r
\end{sequation}
which is equivalent to
\begin{sequation}
   \existsp l_1 \ldots l_q.\
     (\exists y. \bigwedge_{i=1}^q |s_i|=l_i) \ \land\ G_r
\end{sequation}
Without loss of generality assume that the set variables
$s_1,\ldots,s_q$ are numbered such that $s_{2i-1} \equiv
s'_i \cap y^c$ and $s_{2i} \equiv s'_i \cap y$ for some cube $s'_i$.  Then apply
again Lemma~\ref{lemma:mainelim} and replace each pair of conjuncts
\begin{sequation}
  |s'_i \cap y^c|=l_{2i-1} \ \land\ |s'_i \cap y|=l_{2i}
\end{sequation}
with the conjunct $|s'_i| = l_{2i-1} + l_{2i}$, yielding formula
\begin{sequation} \label{eqn:almostThere}
   \existsp l_1 \ldots l_q.\
     \bigwedge_{i=1}^{q'} |s'_i|=l_{2i-1}+l_{2i} \ \land\ G_r
\end{sequation}
for $q' = 2^{e-1}$.  Finally, to obtain a formula of the
form~(\ref{eqn:quantifiedReduction}) for $r+1$, introduce
fresh variables $l'_i$ constrained by $l'_i = l_{2i-1}+l_{2i}$, 
rewrite~(\ref{eqn:almostThere}) as
\begin{sequation*}
   \existsp l'_1 \ldots l'_{q'}.\
     \bigwedge_{i=1}^{q'} |s'_i|=l'_i \ \land\
   (\exists l_1 \ldots l_q.\ \bigwedge_{i=1}^{q'} l'_i=l_{2i-1}+l_{2i} \ \land\ G_r)
\end{sequation*}
and let 
\begin{sequation}
  G_{r+1} \equiv \existsp l_1 \ldots l_q.\ 
   \bigwedge_{i=1}^{q'} l'_i=l_{2i-1}+l_{2i} \ 
   \land\ G_r
\end{sequation}
This completes the description of elimination of an
existential set quantifier $\exists y$.

To eliminate a set quantifier $\forall y$, proceed
analogously: introduce fresh variables $l'_i =
l_{2i-1}+l_{2i}$ and let
$G_{r+1} \equiv \forallp l_1 \ldots l_q.\ 
   (\bigwedge_{i=1}^{q'} l'_i=l_{2i-1}+l_{2i}) \ 
   \implies\ G_r$,
which can be verified by expressing $\forall y$ as $\lnot
\exists y \lnot$. 

After eliminating all quantifiers as described above, we
obtain a formula of the form $\existsp l.\ |\Univ|=l \land
G_{p+1}(l)$.  We define the result of the algorithm, denoted
$\alpha(F_0)$, to be the $\PA$ sentence $G_{p+1}(\MAXCARD)$.

This completes the description of the algorithm $\alpha$.
Given that the validity of $\PA$ sentences is decidable, the
algorithm $\alpha$ is a decision procedure for $\BAPA$
sentences.
\begin{theorem}
  The algorithm $\alpha$ described above maps each
  $\BAPA$-sentence $F_0$ into an equivalent $\PA$-sentence
  $\alpha(F_0)$.
\end{theorem}

\smartparagraph{Formalization of the algorithm $\alpha$.} To
formalize the algorithm $\alpha$, we have implemented it in
the functional programming language O'Caml
(Section~\ref{sec:ocaml}).\footnote{The implementation is
available from \\
\url{http://www.cag.lcs.mit.edu/~vkuncak/artifacts/bapa/}.}
As an illustration, when we run the implementation on the
$\BAPA$ formula in Figure~\ref{fig:exampleVC} which
represents a verification condition, we immediately obtain
the $\PA$ formula in Figure~\ref{fig:translatedVC}.  Note
that the structure of the resulting formula mimics the
structure of the original formula: every set quantifier is
replaced by the corresponding block of quantifiers over
non-negative integers constrained to partition the
previously introduced integer variables.
Figure~\ref{fig:varCorrespondence} presents the
correspondence between the set variables of the $\BAPA$
formula and the integer variables of the translated $\PA$
formula.  Note that the relationship $\content' = \content
\cup e$ translates into the conjunction of the
constraints $|\content' \cap (\content \cup e)^c|=0$ $\land$
$|(\content \cup e) \cap {\content'}^c|=0$, which reduces to
the conjunction $l_{100} = 0 \land l_{011} + l_{001} + l_{010} = 0$
using the translation of set expressions into the disjoint 
union of partitions,
and the correspondence in Figure~\ref{fig:varCorrespondence}.
\begin{figure}
\begin{center}
\begin{minipage}{3in}
\footnotesize
\begin{equation*}
\begin{array}{l}
\forallnat l_{1}. \forallnat l_{0}.\
\MAXCARD = l_{1} + l_{0} \implies \\
  \forallnat l_{11}. \forallnat l_{01}. 
  \forallnat l_{10}. \forallnat l_{00}.\\
  l_{1} = l_{11} + l_{01} \land l_{0} = l_{10} + l_{00} \implies \\
\ \ \forallnat l_{111}.\ \forallnat l_{011}.\
    \forallnat l_{101}.\ \forallnat l_{001}.\\
\ \ 
    \forallnat l_{110}.\ \forallnat l_{010}.\
    \forallnat l_{100}.\ \forallnat l_{000}.\\
  \begin{array}[t]{l}
  \ \
   l_{11} = l_{111} + l_{011} \ \land 
   l_{01} = l_{101} + l_{001} \ \land \\
  \ \
   l_{10} = l_{110} + l_{010} \ \land 
   l_{00} = l_{100} + l_{000} \implies 
  \end{array} \\
\ \ 
\ \ \ \ \ \forallint \mathit{size}. \forallint \mathit{size'}. \\
\ \ 
\ \ \ \ \
\begin{array}[t]{l}
    (l_{111} + l_{011} + l_{101} + l_{001} = 1 \ \land \\
    \> l_{111} + l_{011} = 0 \ \land  \\
    \> l_{111} + l_{011} + l_{110} + l_{010} = \mathit{size} \ \land  \\
    \> l_{100} = 0 \ \land  \\
    \> l_{011} + l_{001} + l_{010} = 0 \ \land  \\
    \> \mathit{size'} = \mathit{size} + 1) \implies \\
\ \ \ \ \ 
    (0 < \mathit{size'} \ \land 
     l_{111} + l_{101} + l_{110} + l_{100} = \mathit{size'})
\end{array}
\end{array}
\end{equation*}
\end{minipage}
\end{center}
\caption{The translation of the $\BAPA$ sentence from Figure~\ref{fig:exampleVC} 
        into a $\PA$ sentence\label{fig:translatedVC}}
\end{figure}

\begin{figure}
\footnotesize
\begin{equation*}
\begin{array}{c}
  \mbox{\bf general relationship:} \\
  l_{i_1,\ldots,i_k} = |\setno_q^{i_1} \cap \setno_{q+1}^{i_2} \cap \ldots \cap \setno_S^{i_k}| \\
  q = S-(k-1), \qquad S - \mbox{number of set variables} \mnl
  \mbox{\bf in this example:} \\
  \begin{array}{r@{\ =\ }l}
    \setno_1 & \content' \\
    \setno_2 & \content \\
    \setno_3 & e \\
  \end{array} \\
  \begin{array}{r@{\ =\ }l}
     l_{000} & |{\content'}^c \cap \content^c \cap e^c| \\
     l_{001} & |{\content'}^c \cap \content^c \cap e| \\
     l_{010} & |{\content'}^c \cap \content \cap e^c| \\
     l_{011} & |{\content'}^c \cap \content \cap e| \\
     l_{100} & |{\content'} \cap \content^c \cap e^c| \\
     l_{101} & |{\content'} \cap \content^c \cap e| \\
     l_{110} & |{\content'} \cap \content \cap e^c| \\
     l_{111} & |{\content'} \cap \content \cap e|
  \end{array}
\end{array}
\end{equation*}
\caption{The Correspondence between Integer Variables in Figure~\ref{fig:translatedVC} and 
        Set Variables in Figure~\ref{fig:exampleVC}\label{fig:varCorrespondence}}
\end{figure}

The subsequent sections explore further consequences of the
existence of the algorithm $\alpha$, including an upper
bound on the computational complexity of $\BAPA$ sentences
and the combination of $\BA$ with proper extensions of
$\PA$.



\section{Complexity}
\label{sec:complexity}

In this section we analyze the algorithm $\alpha$ from
Section~\ref{sec:qeReduction} and obtain space and time
bounds on $\BAPA$ from the corresponding space and time
bounds for $\PA$.  We then show that the new decision
procedure meets the optimal worst-case bounds for Boolean
algebras if applied to purely Boolean algebra formulas.
Moreover, by construction, our procedure reduces to the
procedure for Presburger arithmetic formulas if there are no
set quantifiers.  In summary, our decision procedure is
optimal for $\BA$, does not impose any overhead for pure
$\PA$ formulas, and the complexity of the general $\BAPA$
validity is not much worse than the complexity of $\PA$
itself.

\subsection{An Elementary Upper Bound}
\label{sec:elementary}

We next show that the algorithm in
Section~\ref{sec:qeReduction} transforms a $\BAPA$ sentence
$F_0$ into a $\PA$ sentence whose size is at most one
exponential larger and which has the same number of
quantifier alternations.

If $F$ is a formula in prenex form, let $\size{F}$ denote the size
of $F$, and let $\alts{F}$ denote the number of quantifier
alternations in $F$.  Define the iterated exponentiation
function $\iexp{k}{x}$ by $\iexp{0}{x}=x$ and $\iexp{k+1}{x}
= 2^{\iexp{k}{x}}$.  We have the following lemma.
\begin{lemma} \label{lemma:onlyExp}
  For the algorithm $\alpha$ from
  Section~\ref{sec:qeReduction} there is a constant $c > 0$ such that
  \begin{sequation*}
    \begin{array}{l} 
       \size{\alpha(F_0)} \leq 2^{c \cdot \size{F_0}} \mnl
       \alts{\alpha(F_0)} = \alts{F_0}
    \end{array}
  \end{sequation*}
  Moreover, the algorithm $\alpha$ runs in $2^{O(\size{F_0})}$
  space.
\end{lemma}
\begin{proof}
To gain some intuition on the size of $\alpha(F_0)$ compared
to the size of $F_0$, compare first the formula in
Figure~\ref{fig:translatedVC} with the original formula in
Figure~\ref{fig:exampleVC}.  Let $n$ denote the size of the
initial formula $F_0$ and let $S$ be the number of set
variables.  Note that the following operations are
polynomially bounded in time and space: 1) transforming a
formula into prenex form, 2) transforming relations
$b_1=b_2$ and $b_1 \subseteq b_2$ into the form $|b|=0$.
Introducing set variables for each partition and
replacing each $|b|$ with a sum of integer variables yields
formula $G_1$ whose size is bounded by $O(n 2^S S)$ (the
last $S$ factor is because representing a variable from the
set of $K$ variables requires space $\log K$).  The
subsequent transformations introduce the existing integer
quantifiers, whose size is bounded by $n$, and introduce
additionally $2^{S-1}+\ldots+2+1=2^S-1$ new integer
variables along with the equations that define them.  Note
that the defining equations always have the form $l'_i =
l_{2i-1}+l_{2i}$ and have size bounded by $S$.  We
therefore conclude that the size of $\alpha(F_0)$ is $O(nS
(2^S + 2^S))$ and therefore $O(nS 2^S)$, which is certainly
$O(2^{cn})$ for any $c > 1$.  Moreover, note that we have
obtained a more precise bound $O(nS 2^S)$ indicating
that the exponential explosion is caused only by set
variables.  Finally, the fact that the number of quantifier
alternations is the same in $F_0$ and $\alpha(F_0)$ is
immediate because the algorithm replaces one set quantifier
with a block of corresponding integer quantifiers.
\end{proof}

We next consider the worst-case space bound on $\BAPA$.
Recall first the following bound on space complexity for
$\PA$.
\begin{fact}   \label{fac:FR} 
  \cite[Chapter 3]{FerranteRackoff79ComputationalComplexityLogicalTheories}
  The validity of a $\PA$ sentence of length
  $n$ can be decided in space $\iexp{2}{O(n)}$.
\end{fact}
From Lemma~\ref{lemma:onlyExp} and Fact~\ref{fac:FR} we
conclude that the validity of $\BAPA$ formulas can be
decided in space $\iexp{3}{O(n)}$.  It turns out, however,
that we obtain better bounds on $\BAPA$ validity by
analyzing the number of quantifier alternations in $\BA$ and
$\BAPA$ formulas.
\begin{fact}   \label{fac:RL}
  \cite{ReddyLoveland78PresburgerBoundedAlternation}
  The validity of a $\PA$ sentence of length $n$ and
  the number of quantifier alternations $m$ can be decided in
  space $2^{n^{O(m)}}$.
\end{fact}
From Lemma~\ref{lemma:onlyExp} and Fact~\ref{fac:RL} we
obtain our space upper bound, which implies the upper bound
on deterministic time.
\begin{theorem}
  The validity of a $\BAPA$ sentence of length $n$ and the
  number of quantifier alternations $m$ can be decided in
  space $\iexp{2}{O(mn)}$, and, consequently, in
  deterministic time $\iexp{3}{O(mn)}$.
\end{theorem}
If we approximate quantifier alternations by formula size,
we conclude that $\BAPA$ validity can be decided in space
$\iexp{2}{O(n^2)}$ compared to $\iexp{2}{O(n)}$ bound for
Presburger arithmetic from Fact~\ref{fac:FR}.  Therefore,
despite the exponential explosion in the size of the formula
in the algorithm $\alpha$, thanks to the same number of
quantifier alternations, our bound is not very far from the
bound for Presburger arithmetic.

\subsection{Boolean Algebras as a Special Case}
\label{sec:baspecial}

We next analyze the result of applying the algorithm
$\alpha$ to a pure $\BA$ sentence $F_0$.  By a pure $\BA$
sentence we mean a $\BA$ sentence without cardinality
constraints, containing only the standard operations
$\cap,\cup,{}^c$ and the relations ${\subseteq},{=}$.  At
first, it might seem that the algorithm $\alpha$ is not a
reasonable approach to deciding pure $\BA$ formulas given
that the best upper bounds for $\PA$ are worse than the
corresponding bounds for $\BA$.  However, we identify a
special form of $\PA$ sentences $\PA_{\BA} = \{ \alpha(F_0)
\mid F_0 \mbox{ is in pure } \BA\}$ and show that such
sentences can be decided in $2^{O(n)}$ space, which is
optimal for Boolean algebras
\cite{Kozen80ComplexityBooleanAlgebras}.  Our analysis shows
that using binary representations of integers that
correspond to the sizes of sets achieves a similar effect to
representing these sets as bitvectors, although the two
representations are not identical.



Let $S$ be the
number of set variables in the initial formula $F_0$ (recall
that set variables are the only variables in $F_0$).  Let
$l_1,\ldots,l_q$ be the set of free variables of the formula
$G_r(l_1,\ldots,l_q)$; then $q=2^e$ for $e=S+1-r$.  Let
$w_1,\ldots,w_q$ be integers specifying the values of
$l_1,\ldots,l_q$.  We then have the following lemma.
\begin{lemma} \label{lemma:BAcomplexity}
For each $r$ where $1 \leq r \leq S$ the truth value of
$G_r(w_1,\ldots,w_q)$ is equal to the the truth value of
$G_r(\bar w_1,\ldots,\bar w_q)$ where $\bar w_i =
\min(w_i,2^{r-1})$.
\end{lemma}
\begin{proof}
We prove the claim by induction.  For $r=1$, observe that
the translation of a quantifier-free part of the pure $\BA$
formula yields a $\PA$ formula $F_1$ whose all atomic
formulas are of the form $l_{i_1} + \ldots + l_{i_k}=0$,
which are equivalent to $\bigvee_{j=1}^k l_{i_j}=0$.
Therefore, the truth-value of $F_1$ depends only on whether
the integer variables are zero or non-zero, which means that
we may restrict the variables to interval $[0,1]$.

For the inductive step, consider the elimination of a set
variable, and assume that the property holds for $G_r$ and
for all $q$ tuples of non-negative integers
$w_1,\ldots,w_q$.  Let $q' = q/2$ and $w'_1,\ldots,w'_{q'}$
be a tuple of non-negative integers.  We show that
$G_{r+1}(w'_1,\ldots,w'_{q'})$ is equivalent to
$G_{r+1}(\bar w'_1,\ldots,\bar w'_{q'})$.

Suppose first that $G_{r+1}(\bar w'_1,\ldots,\bar w'_{q'})$
holds.  Then for each $w'_i$ there are $w_{2i-1}$ and
$w_{2i}$ such that $\bar w'_i = u_{2i-1} + u_{2i}$ and
$G_r(u_1,\ldots,u_q)$.  We define witnesses $w_1,\ldots,w_q$
as follows.  If $w'_i \leq 2^r$, then let
$w_{2i-1}=u_{2i-1}$ and $w_{2i}=u_{2i}$.  If $w'_i > 2^r$
then either $u_{2i-1} > 2^{r-1}$ or $u_{2i} > 2^{r-1}$ (or
both).  If $u_{2i-1} > 2^{r-1}$, then let $w_{2i-1} = w'_i -
u_{2i}$ and $w_{2i} = u_{2i}$.  Note that
$G_r(\ldots,w_{2i-1},\ldots) \iff
G_r(\ldots,u_{2i-1},\ldots) \iff G_r(\ldots,2^{r-1},\ldots)$
by induction hypothesis because both $u_{2i-1} > 2^{r-1}$
and $w_{2i-1} > 2^{r-1}$.  For $w_1,\ldots,w_q$ chosen as
above we therefore have $w'_i = w_{2i-1} + w_{2i}$ and
$G_r(w_1,\ldots,w_q)$, which by definition of $G_{r+1}$
means that $G_{r+1}(w'_1,\ldots,w'_{q'})$ holds.

Conversely, suppose that $G_{r+1}(w'_1,\ldots,w'_{q'})$
holds.  Then there are $w_1,\ldots,w_q$ such that
$G_r(w_1,\ldots,w_q)$ and $w'_i = w_{2i-1} + w_{2i}$.  If
$w_{2i-1} \leq 2^{r-1}$ and $w_{2i} \leq w_{2i}$ then $w'_i
\leq 2^r$ so let $u_{2i-1} = w_{2i-1}$ and $u_{2i} =
w_{2i}$.  If $w_{2i-1} > 2^{r-1}$ and $w_{2i} > w_{2i}$ then
let $u_{2i-1} = 2^{r-1}$ and $u_{2i} = 2^{r-1}$.  If
$w_{2i-1} > 2^{r-1}$ and $w_{2i} \leq 2^{r-1}$ then let
$u_{2i-1} = 2^r-w_{2i}$ and $u_{2i} = w_{2i}$.  By induction
hypothesis we have $G_r(u_1,\ldots,u_q) =
G_r(w_1,\ldots,w_q)$.  Furthermore, $u_{2i-1} + u_{2i} =
\bar w'_i$, so $G_{r+1}(\bar w'_1,\ldots,\bar w'_{q'})$ by
definition of $G_{r+1}$.
\end{proof}

Now consider a formula $F_0$ of size $n$ with $S$ free
variables.  Then $\alpha(F_0) = G_{S+1}$.  By
Lemma~\ref{lemma:onlyExp}, $\size{\alpha(F_0)}$ is $O(n S
2^S)$.  By Lemma~\ref{lemma:BAcomplexity}, it suffices for
the outermost variable $k$ to range over the integer
interval $[0,2^S]$, and the range of subsequent variables is
even smaller.  Therefore, the value of each of the
$2^{S+1}-1$ variables can be represented in $O(S)$ space,
which is the same order of space used to represent the names of
variables themselves.  This means that evaluating the
formula $\alpha(F_0)$ can be done in the same space $O(n S
2^S)$ as the size of the formula.  Representing the
valuation assigning values to variables can be done in $O(S
2^S)$ space, so the truth value of the formula can be
evaluated in $O(n S 2^S)$ space, which is
certainly $2^{O(n)}$.  We obtain the following theorem.

\begin{theorem}
  If $F_0$ is a pure $\BA$ formula with $S$ variables and of
  size $n$, then the truth value of $\alpha(B_0)$ can be
  computed in $O(n S 2^S)$ and therefore $2^{O(n)}$
  space.
\end{theorem}



\section{Allowing Infinite Sets}
\label{sec:infSets}

We next sketch the extension of our algorithm $\alpha$
(Section~\ref{sec:qeReduction}) to the case when the
universe of the structure may be infinite, and the
underlying language has the ability to distinguish between
finite and infinite sets.  Infinite sets are useful in
program analysis for modelling pools of objects such as
those arising in dynamic object allocation.

We generalize the language of $\BAPA$ and the interpretation
of $\BAPA$ operations as follows.
\begin{enumerate} \compr
\item Introduce unary predicate $\bafin{b}$ which is 
true iff $b$ is a finite set.  The predicate $\bafin{b}$
allows us to generalize our algorithm to the case
of infinite universe, and additionally gives the expressive
power to distinguish between finite and infinite sets.  For
example, using $\bafin{b}$ we can express bounded
quantification over finite or over infinite sets.
\item Define $|b|$ to be the integer zero if $b$ is 
infinite, and the cardinality of $b$ if $b$ is finite.
\item Introduce propositional variables denoted by letters 
such as $p,q$, and quantification over propositional
variables.  Extend also the underlying $\PA$ formulas with
propositional variables, which is acceptable because a
variable $p$ can be treated as a shorthand for an integer
from $\{0,1\}$ if each use of $p$ as an atomic formula is
interpreted as the atomic formula $(p=1)$.  Our extended algorithm uses the
equivalences $\bafin{b} \miff p$ to represent the finiteness
of sets just as it uses the equations $|b|=l$ to represent
the cardinalities of finite sets.
\item Introduce a propositional constant $\FINU$ such that
$\bafin{\Univ} \miff \FINU$.  This propositional constant
enables equivalence preserving quantifier elimination over
the set of models that includes both models with finite
universe $\Univ$ and the models with infinite universe $\Univ$.
\end{enumerate}
Denote the resulting extended language $\IBAPA$.

The following lemma generalizes 
Lemma~\ref{lemma:mainelim} for the case of equalities.
\begin{lemma} \label{lemma:infmainelim}
  Let $b_1,\ldots,b_n$ be disjoint sets, 
  $l_1,\ldots,l_n,k_1,\ldots,k_n$ be natural numbers, and
  $p_1,\ldots,p_n,q_1,\ldots,q_n$ be propositional values.
  Then the following two statements are equivalent:
  \begin{enumerate}
  \item There exists a set $y$ such that
  \begin{sequation} \label{eqn:infone}
    \bigwedge_{i=1}^n 
        \begin{array}[t]{l}
        |b_i \cap y| = k_i \land (\bafin{b_i \cap y} \miff  p_i)\  \land\\
        |b_i \cap y^c|=l_i \land (\bafin{b_i \cap y^c}\miff q_i)
        \end{array}
  \end{sequation}
  \item \
  \begin{sequation} \label{eqn:inftwo}
    \bigwedge_{i=1}^n 
        \begin{array}[t]{l}
        (p_i \land q_i \implies |b_i| = k_i + l_i) \ \land\\  
        (\bafin{b_i} \miff (p_i \land q_i))
        \end{array}
  \end{sequation}
  \end{enumerate}
\end{lemma}
\begin{proof}
  ($\implies$) Suppose that there exists a set $y$
  satisfying~(\ref{eqn:infone}).  From $b_i = (b_i \cap y)
  \cup (b_i \cap y^c)$, we have $\bafin{b_i} \miff (p_i
  \land q_i)$.  Furthermore, if $p_i$ and $q_i$ hold, then
  both $b_i \cap y$ and $b_i \cap y^c$ are finite so the
  relation $|b_i|=|b_i \cap y|+|b_i \cap y^c|$ holds.

  ($\explies$) Suppose that~(\ref{eqn:inftwo}) holds.
  For each $i$ we choose a subset $y_i \subseteq b_i$, depending
  on the truth values of $p_i$ and $q_i$, as follows.
  \begin{enumerate} \compr
  \item If both $p_i$ and $q_i$ are true, then $\bafin{b_i}$ holds,
  so $b_i$ is finite.  Choose $y_i$ as any subset of $b_i$ with
  $k_i$ elements, which is possible since $b_i$ has $k_i+l_i$ elements.
  \item If $p_i$ does not hold, but $q_i$ holds, then $\bafin{b_i}$ does
  not hold, so $b_i$ is infinite.  Choose $y_i'$ as any finite set with $l_i$
  elements and let $y_i = b_i \setminus y'_i$ be the corresponding cofinite set.
  \item Analogously, if $p_i$ holds, but $q_i$ does not hold, then $b_i$ is infinite;
  choose $y_i$ as any finite subset of $b_i$ with $k_i$ elements.
  \item If $p_i$ and $q_i$ are both false, then $b_i$ is also infinite; every 
  infinite set can be written as a disjoint union of two infinite sets, so let
  $y_i$ be one such set.
  \end{enumerate}
  Let $y = \bigcup_{i=1}^n y_i$.
  As in the proof of Lemma~\ref{lemma:mainelim}, we have
  $b_i \cap y = y_i$ and $b_i \cap y^c = y_i^c$.  By construction of 
  $y_1,\ldots,y_n$ we conclude that~(\ref{eqn:infone}) holds.
\end{proof}

The algorithm $\alpha$ for $\IBAPA$ is analogous to the algorithm
for $\BAPA$.  In each step, the new algorithm maintains a formula
of the form
\begin{sequation*} \label{eqn:infquantifiedReduction}
 \begin{array}{l}
  Q_p v_p \ldots Q_r v_r. \\
  \quad
   \existsp l_1 \ldots l_q.\ \exists p_1 \ldots p_q.\\
  \qquad
        (\bigwedge_{i=1}^q |s_i|=l_i \land (\bafin{s_i}\miff p_i)) \ \land\ G_r
 \end{array}
\end{sequation*}
As in Section~\ref{sec:qeReduction}, the algorithm
eliminates an integer quantifier $\exists k$ by letting
$G_{r+1} = \exists k.G_r$ and eliminates an integer
quantifier $\forall k$ by letting $G_{r+1} = \forall k.G_r$.
Furthermore, just as the algorithm in
Section~\ref{sec:qeReduction} uses
Lemma~\ref{lemma:mainelim} to reduce a set quantifier to
integer quantifiers, the new algorithm uses
Lemma~\ref{lemma:infmainelim} for this purpose.  The algorithm replaces 
\begin{sequation*}
 \begin{array}{l}
   \exists y.\
   \existsp l_1 \ldots l_q.\ \exists p_1 \ldots p_q.\\
  \qquad
        (\bigwedge_{i=1}^q |s_i|=l_i \land (\bafin{s_i}\miff p_i)) \ \land\ G_r
 \end{array}
\end{sequation*}
with
\begin{sequation*}
 \begin{array}{l}
   \existsp l'_1 \ldots l'_{q'}.\ \exists p'_1 \ldots p'_{q'}.\\
  \qquad
        (\bigwedge_{i=1}^{q'} |s'_i|=l'_i \land (\bafin{s'_i}\miff p'_i)) \ \land\ 
          G_{r+1}
 \end{array}
\end{sequation*}
for $q'=q/2$, and
\begin{sequation*}
  \begin{array}{l}
  G_{r+1} \ \equiv \ \existsp l_1 \ldots l_q.\ \exists p_1,\ldots,p_q.\\
  \qquad \quad
  \qquad \begin{array}[t]{l}
     \big(\bigwedge_{i=1}^{q'} 
        \begin{array}[t]{l}
         (p_{2i-1} \land p_{2i} \implies l'_i=l_{2i-1}+l_{2i}) \ \land\\
         (p'_i \miff (p_{2i-1} \land p_{2i}))\big)
        \end{array} \\
     \land\ G_r
      \end{array}
  \end{array}
\end{sequation*}
For the quantifier $\forall y$ the algorithm analogously generates
\begin{sequation*}
  \begin{array}{l}
  G_{r+1} \ \equiv \ \forallp l_1 \ldots l_q.\ \forall p_1,\ldots,p_q.\\
  \qquad \quad
  \qquad \begin{array}[t]{l}
     \big(\bigwedge_{i=1}^{q'} 
        \begin{array}[t]{l}
         (p_{2i-1} \land p_{2i} \implies l'_i=l_{2i-1}+l_{2i}) \ \land\\
         (p'_i \miff (p_{2i-1} \land p_{2i}))\big)
        \end{array} \\
     \implies\ G_r
      \end{array}
  \end{array}
\end{sequation*}

After eliminating all quantifiers, the algorithm obtains a
formula of the form $\existsp l.\exists p.\ |\Univ|=l \land
(\bafin{\Univ} \miff p) \land G_{p+1}(l,p)$.  We define the
result of the algorithm to be the $\PA$ sentence 
$G_{p+1}(\MAXCARD,\FINU)$.

This completes our description of the generalized algorithm
$\alpha$ for $\IBAPA$.  The complexity analysis from
Section~\ref{sec:complexity} also applies to the generalized
version.  We also note that our algorithm yields an
equivalent formula over any family of models.  A sentence is
valid in a set of models iff it is valid on each model.
Therefore, the validity of a $\IBAPA$ sentence $F_0$ is
given by applying to the formula
$\alpha(F_0)(\MAXCARD,\FINU)$ a form of universal quantifier
over all pairs $(\MAXCARD,\FINU)$ that determine the
characteristics of the models in question.  For example, for the
validity over the models with infinite universe we use
$\alpha(F_0)(0,\boolFalse)$, for validity over all finite
models we use $\forall k. \alpha(F_0)(k,\boolTrue)$, and
for the validity over all models we use the $\PA$ formula
\begin{sequation*}
  \alpha(F_0)(0,\boolFalse) \land \forall k. \alpha(F_0)(k,\boolTrue).
\end{sequation*}
We therefore have the following result, which answers a
generalized version of the question left open in
\cite{Zarba04QuantifierEliminationAlgorithmSetCardinality}.
\begin{theorem}
  The algorithm above effectively reduces the validity of
  $\IBAPA$ sentences to the validity of Presburger
  arithmetic formulas with the same number of quantifier
  alternations, and the increase in formula size exponential
  in the number of set variables; the reduction works for
  each of the following: 1) the set of all models, 2) the
  set of models with infinite universe only, and 3) the set
  of all models with finite universe.
\end{theorem}


\section{Relationship with MSOL over Strings}
\label{sec:msol}

The monadic second-order logic (MSOL) over strings is a
decidable logic that can encode Presburger arithmetic
by encoding addition using one successor symbol and
quantification over sets.  This logic therefore
simultaneously supports sets and integers, so it is natural
to examine its relationship with $\BAPA$.  It turns out that
there are two important differences between MSOL over
strings and $\BAPA$:
\begin{enumerate}
\item $\BAPA$ can express relationships of the form $|A|=k$ where
  $A$ is a set variable and $k$ is an integer variable; such
  relation is not definable in MSOL over strings.
\item In MSOL over strings, the sets contain \emph{integers}
as elements, whereas in $\BAPA$ the sets contain
\emph{uninterpreted elements}.
\end{enumerate}
Given these differences, a natural question is to consider
the decidability of an extension of MSOL that allows stating
relations $|A|=k$ where $A$ is a set of integers and $k$ is
an integer variable.  Note that by saying $\exists k. |A|=k
\land |B|=k$ we can express $|A|=|B|$, so we obtain MSOL
with equicardinality constraints.  However, extensions of
MSOL over strings with equicardinality constraints are known
to be undecidable; we review some reductions in
Section~\ref{sec:undmsol}.  Undecidability results such
as these are what perhaps led to the conjecture that $\BAPA$
itself is undecidable \cite[Page
12]{Zarba04QuantifierEliminationAlgorithmSetCardinality}.
In this paper we have shown that $\BAPA$ is, in fact,
decidable and has an elementary decision procedure.
Moreover, we next present a combination of $\BA$ with MSOL
over $n$-successors that is still decidable.

\subsection{Decidability of MSOL with Cardinalities on Uninterpreted Sets}
\label{sec:decmsol}

Consider the multisorted language $\BAMSOL$ defined as
follows. First, $\BAMSOL$ contains all relations of monadic
second-order logic of $n$-successors, whose variables range
over strings over an $n$-ary alphabet and sets of such strings.
Second, $\BAMSOL$ contains sets of uninterpreted elements
and boolean algebra operations on them.  Third, $\BAMSOL$
allows stating relationships of the form $|x|=k$ where
$x$ is a set of uninterpreted elements and $k$ is a string
representing a natural number.  Because all $\PA$ operations
are definable in MSOL of 1-successor, the algorithm $\alpha$
applies in this case as well.  Indeed, the algorithm
$\alpha$ only needs a ``lower bound'' on the expressive
power of the theory of integers that $\BA$ is combined with:
the ability to state constraints of the form
$l'_i=l_{2i-1}+l_{2i}$, and quantification over integers.
Therefore, applying $\alpha$ to a $\BAMSOL$ formula results
in an MSOL formula.  This shows that $\BAMSOL$ is decidable
and can be decided using a combination of algorithm $\alpha$
and tool such as \cite{KlarlundETAL00MONA}.  By
Lemma~\ref{lemma:onlyExp}, the decision procedure for
$\BAMSOL$ based on translation to MSOL has upper bound of
$\iexp{n}{O(n)}$ using a decision procedure such as
\cite{KlarlundETAL00MONA} based on tree automata
\cite{ComonETAL97Tata}.  The corresponding non-elementary
lower bound follows from the lower bound on MSOL itself
\cite{StockmeyerMeyer02CosmologicalLowerBound}.



\section{Related Work}

\smartparagraph{Presburger arithmetic.}
The original result on decidability of Presburger arithmetic
is
\cite{Presburger29UeberVollstaendigkeitSystemsAritmethikZahlen}
(see \cite[Page 24]{Treinen02FirstOrderTheoriesConcreteDomains} for review).
This decision procedure was improved in \cite{Cooper72TheoremProvingArithmetic}
and subsequently in \cite{Oppen73ElementaryBoundsPresburgerArithmetic}.
The best known
bound on formula size is obtained using bounded model property techniques
\cite{FerranteRackoff79ComputationalComplexityLogicalTheories}.
An analysis based on the number of quantifier alternations
is presented in
\cite{ReddyLoveland78PresburgerBoundedAlternation}.
\cite{ChaiebNipkow03GenericPresburger} presents a
proof-generating version of
\cite{Cooper72TheoremProvingArithmetic}.
The omega test as a decision procedure for Presburger arithmetic
is described in
\cite{PughWonnacott95GoingBeyondIntegerProgrammingOmega}.
\cite{Pugh94CountingPresburger} describes how to compute the
number of satisfying assignments to free variables in a
Presburger arithmetic formula, and describes the
applications for computing those numbers for the purpose of
program analysis and optimization.  Some bounds on
quantifier-elimination procedures for Presburger arithmetic
are presented in \cite{Weispfenning97ComplexityPresburger}.
Automata-theoretic \cite{KlarlundETAL00MONA,
BoudetComon96DiophantineEquationsPresburgerArithmeticFiniteAutomata}
and model checking approaches
\cite{GaneshETAL02DecidingPresburgerArithmeticModelChecking, SeshiaBryant04DecidingQuantifierFreePresburgerFormulas}
can also be used to decide Presburger arithmetic and its
fragments.

\smartparagraph{Boolean Algebras.}
The first results on decidability of Boolean algebras are from
\cite{Skolem19Untersuchungen, Loewenheim15UeberMoegligkeitenRelativkalkuel, Tarski49ArithmeticalClassesTypesBooleanAlgebras}, 
\cite[Chapter 4]{Ackermann54SolvableCasesDecisionProblem} and use quantifier elimination,
from which one can derive small model property; \cite{Kozen80ComplexityBooleanAlgebras} 
gives the complexity of the satisfiability problem.
\cite{CantoneETAL01SetTheoryComputing} gives an overview of
several fragments of set theory including theories with
quantifiers but no cardinality constraints and theories with
cardinality constraints but no quantification over sets.

\smartparagraph{Combinations of Decidable Theories.} The
techniques for combining \emph{quantifier-free} theories
\cite{Nelson81TechniquesProgramVerification,
RuessShankar01DeconstructingShostak} and their
generalizations such as
\cite{Zarba04CombinationProblemAutomatedReasoning,
Zarba04CombiningSetsElements} are of great importance for
program verification.  This paper shows a particular
combination result for \emph{quantified formulas}, which add
additional expressive power in writing specifications.
Among the general results for quantified formulas are
the Feferman-Vaught theorem for products
\cite{FefermanVaught59FirstOrderPropertiesProductsAlgebraicSystems},
and term powers
\cite{KuncakRinard03TheoryStructuralSubtyping,
KuncakRinard03StructuralSubtypingNonRecursiveTypesDecidable}.

Our decidability result is closest to
\cite{Zarba04QuantifierEliminationAlgorithmSetCardinality}
which gives a solution for the combination of Presburger
arithmetic with a notion of sets and quantification of
elements, and conjectures that adding the quantification
over sets leads to an undecidable theory.  The results of
this paper prove that the conjecture is false and give an
elementary upper bound on the complexity of the combined
theory.

\smartparagraph{Analyses of Dynamic Data Structures.}
Our new decidability result enables verification tools to
reason about sets and their sizes.  This capability is
particularly important for analyses that handle dynamically
allocated data structures where the number of objects is
statically unbounded
\cite{LamETAL04GeneralizedTypestateCheckingUsingSets,
  LamETAL04ModularPluggableAnalyses,
  KuncakRinard04GeneralizedRecordsRoleLogic,
  YorshETAL04SymbolicallyComputingMostPrecise,
  YahavRamalingam04SeparationHeterogeneousAbstractions,
  Rugina04QuantitativeShapeAnalysis,
  SagivETAL02Parametric}.
Recently, these approaches were extended to 
handle the combinations of the constraints representing
data structure contents and constraints representing
numerical properties of data structures \cite{Rugina04QuantitativeShapeAnalysis,
ChinETAL03ExtendingSizedTypesCollectionAnalysis}.
Our result provides a systematic mechanism for building precise
and predictable versions of such analyses.


\vspace*{-1em}
\section{Conclusion}

Motivated by static analysis and verification of relations
between data structure content and size, we have introduced
the first-order theory of Boolean algebras with Presburger
arithmetic ($\BAPA$), established its decidability,
presented a decision procedure via reduction to Presburger
arithmetic, and showed an elementary upper bound on the
worst-case complexity.  We expect that our decidability
result will play a significant role in verification of
programs 
\cite{Nelson81TechniquesProgramVerification,
  DetlefsETAL98ExtendedStaticChecking,
  FlanaganETAL02ExtendedStaticCheckingJava,
  MannaGroup96STeP}, especially for programs that manipulate
dynamically changing sets of objects
\cite{LamETAL04GeneralizedTypestateCheckingUsingSets,
  LamETAL04ModularPluggableAnalyses,
  KuncakRinard04GeneralizedRecordsRoleLogic,
  YorshETAL04SymbolicallyComputingMostPrecise,
  YahavRamalingam04SeparationHeterogeneousAbstractions,
  Rugina04QuantitativeShapeAnalysis,
  SagivETAL02Parametric}.


\paragraph{Acknowledgements.}  The first author would
like to thank the members of the Stanford REACT group and
the members of the Berkeley CHESS group on useful
discussions on decision procedures and program analysis, and
Bruno Courcelle on remarks regarding undecidability
of MSOL with equicardinality constraints.



{\footnotesize
\bibliographystyle{plain}
\bibliography{pnew}
}

\section{Appendix}
\subsection{Quantifier Elimination for $\PA$}
\label{sec:qePresburger}

For completeness, this section reviews a procedure for
quantifier elimination in Presburger arithmetic.  For
expository purposes we present a version of the quantifier
elimination procedure that first transforms the formula into
disjunctive normal form.  The transformation to disjunctive
normal form can be avoided, as observed in
\cite{Cooper72TheoremProvingArithmetic, Oppen73ElementaryBoundsPresburgerArithmetic,
ReddyLoveland78PresburgerBoundedAlternation}.  However, our
results in Section~\ref{sec:bapa} can be used with other
variations of the quantifier-elimination for Presburger
arithmetic, and can be formulated in such a way that they
not only do not depend on the technique for quantifier
elimination for Presburger arithmetic, but do not depend on
the technique for deciding Presburger arithmetic at all,
allowing the use of automata-theoretic 
\cite{KlarlundETAL00MONA} and model checking techniques
\cite{GaneshETAL02DecidingPresburgerArithmeticModelChecking}.

Figure~\ref{fig:PresburgerSyntax} presents the syntax of
Presburger arithmetic formulas.  We interpret formulas over
the structure of integers, with the standard interpretation
of logical connectives, quantifiers, irreflexive total order
on integers, addition, subtraction, and constants.  We allow
multiplication by a constant only (the case $C \cdot T$ in
Figure~\ref{fig:PresburgerSyntax}), which is expressible
using addition and subtraction.  If $c$ is a constant and
$t$ is a term, the notation $\pdvd{c}{t}$ denotes that $c$
divides $t$ i.e., $t \bmod c = 0$.  We assume that $c > 0$
in each formula $\pdvd{c}{t}$.

\begin{figure}
  \begin{sequation*}
    \begin{array}{rcll}
      F & ::= & A \mid F_1 \land F_2 \mid F_1 \lor F_2 \mid \lnot F \mid
                \exists x.F \mid \forall x.F \mnl
      A & ::= & T_1 = T_2 \mid T_1 < T_2 \mid \pdvd{C}{T} \mnl
      T & ::= & C \mid T_1 + T_2 \mid T_1 - T_2 \mid C \cdot T \mnl
      C & ::= & \ldots {-2} \mid -1 \mid 0 \mid 1 \mid 2 \ldots
    \end{array}
  \end{sequation*}
  \caption{Formulas of Presburger Arithmetic $\PA$
    \label{fig:PresburgerSyntax}}
\end{figure}

We review a simple algorithm for deciding Presburger
arithmetic inspired by
\cite{Presburger29UeberVollstaendigkeitSystemsAritmethikZahlen},
\cite[Page 24]{Treinen02FirstOrderTheoriesConcreteDomains},
\cite{Cooper72TheoremProvingArithmetic}.  The algorithm we
present eliminates an existential quantifier from a
conjunction of literals in the language of
Figure~\ref{fig:PresburgerSyntax}, which suffices by 
Section~\ref{sec:qeIntro}.  Note first that we may eliminate
all equalities $t_1=t_2$ because
\begin{sequation*}
  t_1 = t_2 \ \iff\ (t_1 < t_2 + 1) \lor (t_2 < t_1 + 1)
\end{sequation*}
Next, we have $\lnot (t_1 < t_2) \ \iff\ t_2 < t_1 + 1$
and
\begin{sequation*}
  \lnot (\pdvd{c}{t}) \ \iff\ \bigvee_{i=1}^{c-1} \pdvd{c}{t{+}i}
\end{sequation*}
which means that it suffices to consider the elimination of
an existential quantifier from the formula of the form
$\bigwedge_{i=1}^n A$ where each $A$ is an atomic formula of
the form $t_1 < t_2$ or of the form $\pdvd{c}{t}$.  Each
of the terms $t_1,t_2,t$ is linear, so we can write it in the form $c_0 +
\sum_{i=1}^k c_i x_i$.  Consequently, we may transform the
atomic formulas into forms $0 < c_0 + \sum_{i=1}^k c_i x_i$
and $\pdvd{c}{c_0 + \sum_{i=1}^k c_i x_i}$.  Consider an
elimination of an existential quantifier $\exists x$ from a
conjunction of such atomic formulas.  Let $c_1,\ldots,c_p$
be the coefficients next to $x$ in the conjuncts and let $M
> 0$ be the least common multiple of $c_1,\ldots,c_p$.
Multiply each atomic formula of the form $0 < c_i x + t$ by
$M/|c_i|$, and multiply each atomic formula of the form
$\pdvd{c}{c_i x + t}$ by $M/c_i$ (yielding $\pdvd{M c}{M x +
(M/c_i)t}$).  The result is an equivalent conjunction of
formulas with the property that, in each conjunct, the
coefficient next to $x$ is $M$ or $-M$.  The conjunction is
therefore of the form $F_0(M x)$ for some formula $F_0$.
The formula $\exists x.F_0(M x)$ is equivalent to the
formula $\exists y. (F_0(y) \land \pdvd{M}{y})$.  By moving
$x$ to the left-hand side if its coefficient is $-1$ in the term
$t$ of each atomic formula $0 < t$, replacing
$\pdvd{c}{-y+t}$ by $\pdvd{c}{y-t}$, and renaming $y$ as
$x$, it remains to eliminate an existential quantifier from
$\exists x. F(x)$ where
\begin{sequation*}
   F(x) \ \equiv \
    \bigwedge_{i=1}^q x < a_i \ \land \
    \bigwedge_{i=1}^p b_i < x \ \land \
    \bigwedge_{i=1}^r \pdvd{c_i}{x+t_i}
\end{sequation*}
where $x$ does not occur in any of $a_i$, $b_i$, $t_i$.  Let
$N$ be the least common multiple of $c_1,\ldots,c_r$.
Clearly, if $x=u$ is a solution of $F_1(x) \equiv
\bigwedge_{i=1}^r \pdvd{c_i}{x+t_i}$, then so is $x = u + Nk$
for every integer $k$.
If $p=0$ and $q=0$ then $\exists y. F(y)$ is equivalent to
e.g.\ $\bigwedge_{i=1}^N F(i)$,
which eliminates the quantifier.  Otherwise, suppose that $p
> 0$ (the case $q > 0$ is analogous, and if $p > 0$ and $q >
0$ then both are applicable).  Suppose for a moment that we
are given an assignment to free variables of $\exists
x.F(x)$.  Then the formula $\exists x.F(x)$ is equivalent to
$\bigvee_u F_1(u)$ where $u$ ranges over the elements $u$
such that
\begin{sequation*}
  \max(b_1,\ldots,b_p) < u < \min(a_1,\ldots,a_q)
\end{sequation*}
Let $b = \max(b_1,\ldots,b_p)$.  Then $\exists x.F(x)$ is
equivalent to $\bigvee_{i=1}^N F(b+i)$.  Namely, if a
solution exists, it must be of the form $b+i$ for some $i >
0$, and it suffices to check $N$ consecutive numbers as
argued above.  Of course, we do not know the assignment to
free variables of $\exists x.F(x)$, so we do not know for
which $b_i$ we have $b=b_i$.  However, we can check all
possibilities for $b_i$.  We therefore have that $\exists
y.F(y)$ is equivalent to
\begin{sequation*}
  \bigvee_{j=1}^p \bigvee_{i=1}^N F(b_j + i)
\end{sequation*}
This completes the sketch of the quantifier elimination for
Presburger arithmetic.  We obtain the following result.

\begin{fact} 
  For every first-order formula $\phi$ in the language of
  Presburger arithmetic of Figure~\ref{fig:PresburgerSyntax}
  there exists a quantifier-free formula $\psi$ such that
  $\psi$ is a disjunction of conjunctions of literals, the
  free variables of $\psi$ are a subset of the free
  variables of $\phi$, and $\psi$ is equivalent to $\phi$
  over the structure of integers.
\end{fact}



\subsection{Undecidability of MSOL of Integer Sets with Cardinalities}
\label{sec:undmsol}

We first note that there is a reduction from the Post
Correspondence Problem that shows the undecidability of MSOL
with equicardinality constraints.  Namely, we can represent
binary strings by finite sets of natural numbers.  In this
encoding, given a position, MSOL itself can easily express
the local property that, at a given position, a string
contains a given finite substring.  The equicardinality
gives the additional ability of finding an $n$-th element of
an increasing sequence of elements.  To encode a PCP
instance, it suffices to write a formula checking the
existence of a string (represented as set $A$) and the
existence of two increasing sequences of equal length
(represented by sets $U$ and $D$), such that for each $i$,
there exists a pair $(a_j,b_j)$ of PCP instance such that
the position starting at $U_i$ contains the constant string
$a_j$, and $U_{i+1} = U_i + |a_j|$, and similarly the
position starting at $D_i$ contains $b_j$ and $D_{i+1} = D_i
+ |b_j|$.

The undecidability of MSOL over strings extended with
equicardinality can also be shown by encoding multiplication
of natural numbers.  Given $A = \{1,2,...,x\}$ and $B =
\{1,2,...,y\}$, we can define a set the set $C =
\{x,2x,...y\cdot x\}$ as the set with the same number of
elements as $B$, that contains $x$, and that is closed under
unary operation $z \mapsto z+y$.  Therefore, if we represent
a natural number $n$ as the set $\{1,\ldots,n\}$, we can
define both multiplication and addition of integers.  This
means that we can write formulas whose satisfiability
answers the existence of solutions of Diophantine equations,
which is undecidable by
\cite{Matiyasevich70EnumerableDiophantine}.  A similar
reduction to a logic that does not even have quantification
over sets is presented in
\cite{Zarba04QuantifierEliminationAlgorithmSetCardinality}.


\newpage
\subsection{O'Caml source code of algorithm $\alpha$}
\label{sec:ocaml}
{\tiny
\verbatiminput{alpha.txt}}

\end{document}